\documentclass[12pt]{amsart}
\usepackage{amsbsy,amssymb,amscd,amsfonts,latexsym,amstext,delarray,
amsmath,graphicx} 
\input xypic

\usepackage{color}

\newtheorem{thm}{Theorem}[section]
\newtheorem{prop}[thm]{Proposition}

\newtheorem{lem}[thm]{Lemma}

\newtheorem{defn}[thm]{Definition}

\newtheorem{rem}[thm]{Remark}
\newtheorem{ex}[thm]{Example}

\numberwithin{equation}{section}

\def\bF{{\mathbb F}}
\def\bG{{\mathbb G}}

\def\bK{{\mathbb K}}
\def\bL{{\mathbb L}}

\def\bT{{\mathbb T}}

\def\A{{\mathbb A}}
\def\C{{\mathbb C}}
\def\F{{\mathbb F}}
\renewcommand{\H}{{\mathbb H}}
\def\N{{\mathbb N}}
\renewcommand{\P}{{\mathbb P}}
\def\Q{{\mathbb Q}}
\def\R{{\mathbb R}}
\def\Z{{\mathbb Z}}
\def\K{{\mathbb K}}

\def\cA{{\mathcal A}}
\def\cB{{\mathcal B}}

\def\cE{{\mathcal E}}
\def\cF{{\mathcal F}}

\def\cH{{\mathcal H}}
\def\cI{{\mathcal I}}
\def\cJ{{\mathcal J}}

\def\cL{{\mathcal L}}

\def\cN{{\mathcal N}}
\def\cO{{\mathcal O}}
\def\cP{{\mathcal P}}

\def\cV{{\mathcal V}}

\def\GL{{\rm GL}}
 
\def\Hom{{\rm Hom}}

\def\Spec{{\rm Spec}}

\def\Tr{{\rm Tr}}

\def\fa{{\mathfrak a}}

\def\fh{{\mathfrak h}}

\def\fH{{\mathfrak H}}

\title[Functor of Points, Heights, Noncommutative Arakelov]{Functor of Points and Height Functions for Noncommutative Arakelov Geometry}
\author{Alicia Lima \& Matilde Marcolli}

\address{Perimeter Institute for Theoretical Physics, Waterloo \\ Canada \newline \indent
Department of Mathematics, The University of Chicago, Chicago, IL \\ USA}
\email{alima@uchicago.edu}
\address{Mathematics Department, California Institute of Technology, Pasadena \\ USA}
\email{matilde@caltech.edu}

\begin{document}
\maketitle

\begin{abstract}
We propose a notion of functor of points for noncommutative spaces, valued in categories of bimodules,
and endowed with an action functional determined by a notion of hermitian structures and height functions,
modeled on an interpretation of the classical functor of points as a physical sigma model.
We discuss different choices of such height functions, based on different notions of volumes and traces, 
including one based on the Hattori-Stallings rank. We show that the height function determines a
dynamical time evolution on an algebra of observables associated to our functor of points. 
We focus in particular the case of noncommutative
arithmetic curves, where the relevant algebras are sums of matrix algebras over division algebras over
number fields, and we discuss a more general notion of noncommutative arithmetic spaces in higher
dimensions, where our approach suggests an interpretation of the Jones index as a height function.
\end{abstract}

\section{Introduction}

The use of noncommutative geometry methods in Arakelov geometry was originally
introduced in \cite{ConsMar1}, \cite{ConsMar2}, with a description of the special fiber
at infinity of arithmetic surfaces in terms of a noncommutative space related to the
geometry of Schottky uniformization of Riemann surfaces. A formalism for Arakelov
geometry for noncommutative arithmetic curves and surfaces, based on arithmetic
vector bundles and height functions, was later developed in \cite{Borek1}, \cite{Borek2}. 
This second approach is based on noncommutative projective algebraic geometry in
the sense of Artin--Zhang \cite{Artin}, 
while the results of \cite{ConsMar1}, \cite{ConsMar2} used noncommutative
differential geometry in the sense of Connes. In particular, the use of \cite{Artin}
limits the approach of \cite{Borek1}, \cite{Borek2} to the cases of arithmetic
curves and surfaces. The question of whether there is a good way to 
connect these two approaches and include higher dimensional arithmetic
noncommutative spaces is mentioned in \cite{Borek2}. One of the goals
of the present paper is to establish a formalism that adapts the viewpoint of
\cite{Borek1} and \cite{Borek2} to work in a setting compatible with 
noncommutative geometry in the sense of Connes and covering arithmetic
noncommutative spaces of arbitrary dimension.

\smallskip

Our formalism to describe arithmetic noncommutative spaces
is based on the following main ideas:
\begin{itemize}
\item It is possible to have a good notion of {\em functor of points} for noncommutative
spaces, provided the usual set-valued notion of $S$-points of a scheme $X=\Spec(R)$
given by $X(S)=\Hom_{{\rm Alg}}(R,S)$ is replaced by a functor with values
in categories, that replaces the set $\Hom_{{\rm Alg}}(R,S)$ with the category of
$R$-$S$ bimodules. 
\item The dual notion of functor of points as $\pi_X: {\rm Sch}^{op}\to {\rm Sets}$ 
with $\pi_X(Y)=\Hom_{{\rm Sch}}(Y,X)$ can be envisioned as a $\sigma$-model
with target $X$, where the points, given by maps $\phi\in \Hom_{{\rm Sch}}(Y,X)$, 
are weighted by an action functional measuring their energy, in the form of a {\em height function}.
\item Suitable height functions can be defined for bimodules, and they give rise to a
dynamical system on a convolution algebra of bimodules with the tensor product operation,
with a partition function that corresponds to a height zeta function.
\item The Jones index of Hilbert $C^*$-bimodules of finite type can be seen as a 
possible height function for noncommutative spaces. 
\end{itemize}

\smallskip

In recent years extensions of $\sigma$-models to noncommutative spaces have been studied
in various contexts, sometime motivated by string theory (see for instance \cite{Kap}, \cite{MathRos}),
sometimes by extending results on the geometry of solitons to the noncommutative framework
(see for instance \cite{DaLaLu}). As shown in \cite{DaKraLa}, \cite{DaKraLa2}, several interesting
new phenomena occur in the noncommutative setting and even very simple cases, such as the
example of a target space consisting of a two-point space can have highly nontrivial solutions. 

\smallskip

Here we take a different viewpoint. We broadly regard $\sigma$-models
as describing dynamics on a space of maps between two assigned geometries. Instead of the
usual Riemannian viewpoint, however, we start from a different commutative setting,
consisting of the functor of points of algebraic geometry.

\smallskip

In order to make the functor of points ``dynamical", we need to assign a suitable ``action
functional" to points of an affine scheme. We argue that a good measure of the ``energy"
of points (seen as maps between affine schemes) is provided by the {\em height} function. 
The idea that height functions should be regarded as a physical action functional, with the
height zeta function playing the role of the partition function of the system, was already
suggested by Manin in \cite{Manin95}.

\smallskip

We adapt this viewpoint to the noncommutative setting. It is well known that an analogous notion
of functor of points for noncommutative rings is problematic, due to the scarcity of two-sided
ideals, hence of non-trivial ring homomorphisms (points in the classical sense). It is also well known, however, that
in noncommutative geometry morphisms of algebras are not the most natural choice of
morphisms of noncommutative spaces, and bimodules are a much more natural choice,
in view of phenomena such as Morita equivalence, which provides the appropriate notion
of isomorphism for noncommutative spaces. Thus, we propose replacing the functor of
points of affine schemes with a functor from the category of (noncommutative) rings
with morphisms given by bimodules, with values not in the category of sets, but in the
2-category of small categories. Under this noncommutative ``functor of points"  the
``$S$-points" of a ring $R$ form the category of $R-S$ bimodules.  

\smallskip

We then seek an appropriate replacement for the height function that makes 
this notion of points dynamical. We present two different proposals for a height
function on bimodules. The first is based on extending to our setting a notion
of height developed for noncommutative arithmetic curves in \cite{Borek1}
and based on volumes associated to hermitian forms on bimodules. We show
that this notion of height carries over to our setting but it has the drawback that
it is not always compatible with the composition operation given by the
tensor product of bimodules. The compatibility can be restored by restricting to
a suitable subcategory of bimodules. The second form of height function that
we consider is naturally compatible with tensor products of bimodules and is
based on the Hattori--Stallings rank. We show that, when generalized from 
noncommutative arithmetic curves to higher dimensional noncommutative 
arithmetic spaces, this second notion gives rise to an interpretation of the
Jones index as a height function. 

\smallskip

The paper is organized in the following way. In \S \ref{PtsHSec} we
review some known facts about height functions that serve as
background motivation for what follows and justify our thinking of
the height as an action functional. In \S \ref{BimodPtsSec} we
introduce our proposal for a functor of points in noncommutative
geometry that uses categories of bimodules instead of sets of
ring homomorphisms. In \S \ref{FinDimSec} we focus on the
case of finite-dimensional semisimple algebras, and correspondingly
of noncommutative arithmetic curves. We show that the height
function used in \cite{Borek1} can be generalized to our setting,
but is only compatible with tensor product of bimodules if we
restrict it to a specific subcategory. We propose an alternative
notion of height, based on the Hattori--Stallings rank, which is
better behaved under tensor product of bimodules and we
discuss the time evolutions generated by these height functions
seen as energy functionals. In \S \ref{higherdimSec} we 
introduce a setting describing arithmetic noncommutative
spaces in higher dimension and their functor of points,
using hermitian bimodules of finite type. We consider
as an example, arithmetic structures on noncommutative tori.
We show that the two notions of height discussed in the
case of finite dimensional algebras extend to this setting.
The natural generalization of the height based on the
Hattori--Stallings rank uses the Jones index of Hilbert
$C^*$-bimodules.

\medskip

\section{Functor of Points and Height Dynamics} \label{PtsHSec}

Let $X$ be an affine scheme over a field $\K$. We will assume that $\K$ is
number field. In fact, for our purposes we can just take $\K=\Q$. Points of
the scheme $X$ are defined through the {\em functor of points}, from the
opposite category of schemes to sets
\begin{equation}\label{hX}
\pi_X: {\rm Sch}^{op} \to {\rm Sets}, \ \ \  \pi_X(Y) =\Hom_{{\rm Sch}}(Y,X).
\end{equation}
The set $\pi_X(Y)$ describes the $Y$-points of $X$. An equivalent way of
formulating the functor of points is dually in terms of the corresponding 
algebras. Let $R$ be a commutative $\K$-algebra with $X={\rm Spec}(R)$. 
We then consider
\begin{equation}\label{hR}
\pi^\vee_X: {\rm Alg}_\K \to {\rm Sets}, \ \ \  \pi^\vee_X(S) =\Hom_{{\rm Alg}_\K}(R,S),
\end{equation}
where ${\rm Alg}_\K$ is the category of commutative algebras over $\K$.
The set $\pi^\vee_X(S)$ is the set of $S$-points of $X$. This includes
for example, the case of algebraic points of $X$, that is $X(\bar\Q):=\pi^\vee_X(\bar\Q)$.
It is customary to simply use the notation $X(S)$ for the set $\pi^\vee_X(S)$ of 
$S$-points of the scheme $X$, for a $\K$-algebra $S$ and we will adopt this
notation too. 

\smallskip

More concretely, one can think of an affine scheme $X$ over the field $\K$ as 
solutions to a set of polynomial equations $f_1(x_1,\ldots, x_n)=0, \ldots, f_N(x_1,\ldots, x_n)=0$
for a given set of polynomials in $\K[x_1,\ldots,x_n]$. If we identify the scheme with the system of
equations, then the set $X(S)$ of $S$-points of $X$ for a $\K$-algebra $S$ represents the
set of solutions to these polynomial equations with $x_i \in S$. (The set of $n$-tuples of
elements of the algebra $S$ that satisfy the system of polynomial equations.) This description of
the scheme $X$ is clearly equivalent to describing it through the ring $R=\K[x_1,\ldots,x_n]/\cI_X$
with the ideal $\cI_X=(f_1, \ldots, f_N)$. Any solution with $x_i \in S$ defines a morphism
$R \to S$, hence we recover the previous description of the functor of points. Thinking in terms
of systems of equations, we then describe morphisms of affine schemes as polynomial transformations
between the sets of solutions of two different systems of equations. In other words, if $Y$ is determined
by equations $g_1(y_1,\ldots, y_r)=0, \ldots, g_L(y_1,\ldots, y_r)=0$ and $X$ by equations 
$f_1(x_1,\ldots, x_n)=0, \ldots, f_N(x_1,\ldots, x_n)=0$, a morphism $\sigma: Y \to X$ can be
described as a set of polynomials $\ell_1(y_1,\ldots,y_r), \ldots, \ell_n(y_1,\ldots, y_r)$ such that,
if $(s_1,\ldots, s_r)$ with $s_i\in S$ is a solution to the first system, a point in $Y(S)$, then
$\ell_1(s_1,\ldots, s_r), \ldots, \ell_n(s_1,\ldots, s_r) \in S$ give a solution to the second system,
a point in $X(S)$. Equivalently the substitution of variables 
$(x_1=\ell_1(y_1,\ldots,y_r), \ldots, x_n=\ell_n(y_1,\ldots, y_r))$ defines
a morphism $\K[x_1,\ldots,x_n]\to \K[y_1,\ldots,y_r]$ that descends to
the quotients $R_X=\K[x_1,\ldots,x_n]/\cI_X \to R_Y=\K[y_1,\ldots,y_r]/\cI_Y$ with
$\cI_X=(f_1,\ldots,f_N)$ and $\cI_Y=(g_1,\ldots,g_L)$. We spelled out morphisms
explicitly in this way, since this will be useful in the next subsection. 

\smallskip

A scheme is determined up to isomorphism by its functor of points. 
For a general introduction to the geometry of schemes we refer the
reader to \cite{Man}.

\smallskip
\subsection{Height as an Action Functional}

We consider here the functor of points as providing the kinematics,
namely the space of maps to the target space given by the scheme 
$X$ from a (variable) source space given by the scheme $Y$. In order
to make this model dynamical, we need to specify an action functional
with respect to which the maps $\sigma \in \Hom_{{\rm Sch}}(Y, X)$ are
weighted. 

\smallskip

We recall a few facts about the height function. We refer the reader to \cite{BoGu} for a detailed introduction.
As above let $\K$ be a number field with $d=[\K:\Q]$. Let $\cP_\K$ be the set of places
of $\K$, with $\cP_\K=\cP_\K^{ar} \cup \cP_\K^{nar}$ where $\cP_\K^{ar}$ is the
set of archimedean places, consisting of $r$ real embeddings $\K \hookrightarrow \R$
and $k$ conjugate pairs of complex embeddings $\K \hookrightarrow \C$ (not contained in $\R$) 
with $d=r+2k$, and $\cP_\K^{nar}$ is the set of non-archimedean places. For $\nu\in \cP_\K$ we
write $\K_\nu$ for the corresponding local field (an extension of a $p$-adic field
$\Q_\nu$ at the non-archimedean places and a copy of $\R$ or $\C$ at the archimedean ones).
Let $d_\nu=[ \K_\nu : \Q_\nu ]$ be the degree of the extension at the non-archimedean places, 
and equal to either $1$ or $2$ at the archimedean cases that are either real or complex. 
We also write $| x |_\nu$ for the corresponding absolute values normalized so as to satisfy
the product formula $\prod_{\nu\in \cP_\K} | x |_\nu^{d_\nu}=1$. 

\smallskip

The height of an algebraic number $\alpha \in \bar\Q$ is defined as
\begin{equation}\label{Halpha}
H(\alpha)=\prod_{\nu\in \cP_\K} \max \{ 1, |\alpha|_\nu \}^{d_\nu/d},
\end{equation}
and the logarithmic height as
\begin{equation}\label{halpha}
h(\alpha) =\log H(\alpha) =\sum_{\nu\in \cP_\K} \frac{d_\nu}{d} \log^+ |\alpha|_\nu
\end{equation}
with $\log^+ t=\log \max\{ 1,t \}$. Here $d$ and $d_\nu$ and the set of places $\cP_\K$ are 
taken with respect to a choice of a finite extension $\K$ of $\Q$ that contains $\alpha$, though the resulting
value is independent of this choice. The height of an algebraic number satisfies $h(\alpha)=0$ iff $\alpha$ is a
root of unity. 

\smallskip

In a similar way, one can define the height of a polynomial. One defines for $$f(x_1,\ldots, x_n)=\sum_{k_1,\ldots k_n}
\alpha_{k_1,\ldots, k_n} \, x_1^{k_1}\cdots x_n^{k_n} \in \K[x_1,\ldots, x_n] $$ 
\begin{equation}\label{heightf}
h(f)= \sum_{\nu\in \cP_\K} \frac{d_\nu}{d} \log |f|_\nu  \, , \ \ \ \text{ where } \ \ \ |f|_\nu=\max_{k_1,\ldots,k_n} | \alpha_{k_1,\ldots, k_n} |_\nu \, .
\end{equation}
In fact, it is customary to replace in the archimedean contribution to the height $h(f)$ in the above
formula the term $|f|_\nu$ with the better behaved {\em Mahler measure}
\begin{equation}\label{Mf}
M(f) =\exp \left(\frac{1}{(2\pi)^n} \int_{\bT^n} \log | f(e^{i\theta_1}, \ldots, e^{i\theta_n}) |\, d\theta_1\cdots d\theta_n \right)\, .
\end{equation}
Indeed, while at the non-archimedean places one has $|fg|_\nu = |f|_\nu \cdot |g|_\nu$ this
multiplicativity does not hold at the archimedean places, while the Mahler measure satisfies $M(fg)=M(f) M(g)$.

\smallskip

Given two affine schemes $X$ and $Y$ and a morphism $\sigma \in \Hom_{{\rm Sch}}(Y,X)$,
we can then define a height of $\sigma$, by identifying as in the previous section the morphism
$\sigma$ with a collection of polynomials $\ell_1(y_1,\ldots,y_r), \ldots, \ell_n(y_1,\ldots,y_r)$
that transform by change of variables
$S$-solutions of the polynomial equations defining $Y$ to $S$-solutions of the polynomial 
equations defining $X$. We then define, for the vector $\ell=( \ell_i )_{i=1}^n$ in $\K[y_1,\ldots, y_r]$
the quantity $\| \ell \|_\nu =\max_{i=1,\ldots,n} |\ell_i|_\nu$ and we compute the height
$h(\sigma)$ as in \eqref{heightf}, \eqref{Mf} using this quantity (with $M(\ell_i)$ instead of $|\ell_i|_\nu$
at the archimedean places). 

\smallskip

The definition above provides us with a real valued action functional
\begin{equation}\label{acth}
h: \Hom_{{\rm Sch}}(Y,X) \to \R, 
\end{equation}
with $h(\sigma)$ given by the height of the morphism $\sigma$.
This means that we are thinking of the height as measuring the ``energy"
of the points in $X(S)$ for $Y={\rm Spec}(S)$.

\smallskip

It is reasonable to think of the height as an energy functional. For
example, the Mahler measure determines the free energy and the
growth rate of BPS states in  toric quiver gauge theories, \cite{Zah}.
Mahler measure of certain bivariate polynomials arises as the free 
energy in the planar dimer model, \cite{KeOkShe}, 
for Mahler measures and dimer models see also \cite{Stie1}.
The Mahler measure is also related to instanton expansion, \cite{Stie2}.
The height function can be viewed as a measure of arithmetic complexity,
hence minimizing an action functional based on height means selecting
points of minimal arithmetic complexity. 

\smallskip
\subsection{The minimization problem}

With our choice of energy functional \eqref{acth} then looking for
energy minimizers means minimizing the height function over systems
of polynomials $\ell=(\ell_i)_{i=1}^n: \A^r \to \A^n$, subject to the 
constraints imposed by the source and target schemes $Y$ and $X$,
namely that the polynomial map $\ell$ descends to a map of the
quotients $\ell: \K[x_1,\ldots, x_n]/\cI_X \to \K[y_1,\ldots, y_r]/\cI_Y$.

\smallskip

However, it is clear by looking at the archimedean contribution to 
the action functional given by the Mahler measure that if we want to
obtain a non-trivial and interesting minimization problem for this
energy functional, we need to be more restrictive in the choice of
our class of maps. Indeed, it is better to work under the assumption
that our schemes are defined over $\Z$ (or the ring of integers of
a number fields). Let us assume for this discussion that they
are defined over $\Z$. This means that we only need to consider
polynomials with $\Z$ coefficients, both in the defining equations
of our schemes and in the morphisms between them,
$$ \ell: \Z[x_1,\ldots, x_n]/\cI_X \to \Z[y_1,\ldots, y_r]/\cI_Y. $$
We are then seeking to minimize the height over such polynomials.
That this is now an interesting minimization problem can be seen 
already in the case where both $X=Y=\A^1$ and a single
polynomial map $\ell: \A^1 \to \A^1$. In this case the Mahler 
measure $M(\ell)\geq 1$ and it is known that equality holds 
whenever $\ell(x)$ is a product of cyclotomic polynomials and
the monomial $x$. So these realize the minima of the Mahler
measure. More interesting is the question of what are the polynomials
with the smallest value of the Mahler measure that is $M(\ell)>1$.
This is known as the Lehmer problem, see \cite{Hiro}, \cite{GhaHiro}, \cite{Moss}.
Lehmer noted that the polynomial $\ell(x)=x^{10}+x^9-x^7-x^6-x^5-x^4-x^3+x+1$
has $M(\ell)=1.1762808\ldots$ and the question of whether this is an
optimal lower bound is an open question. This shows that indeed there
are interesting minimization questions associated to the action
functional we are considering. 

\smallskip

It is also worth pointing out the following aspect of this problem.
Consider a scheme $X$ and its set of rational points $X(\K)$.
It is known that the distribution of rational points of bounded height 
is very not uniform: these points accumulate on subsets that are the 
range of certain morphisms. In this case then the question becomes
the search for algebraic points of minimal height (for the appropriate
version of the height function).

\smallskip
\subsection{The case of correspondences}

This commutative case is discussed here only as motivation, hence
we do not discuss in further detail this minimization problem. We do,
however, discuss a variant of it, where morphisms $\sigma: Y \to X$
are replaced by more general correspondences, given by subschemes 
$\Gamma \subset Y\times X$. The case of a morphism $\sigma: Y \to X$
is recovered as the particular case where $\Gamma_\sigma \subset Y\times X$
is the graph of the morphism $\Gamma_\sigma=\{ (y,\sigma(y))\in Y\times X \,|\, y\in Y \}$.
Passing from morphisms given by maps of algebraic varieties or schemes to 
morphisms given by correspondences is a very natural step in algebraic
geometry and it is crucial in the development of the theory of motives.
For us, the reason why it is useful to extend the setting discussed above from maps to
correspondences lies in the fact that morphisms given by correspondences are the
commutative analog of morphisms given by bimodules in noncommutative geometry
(see a discussion of this in Chapter~4 of \cite{CoMa}). Passing from morphisms
of varieties to correspondences is very delicate, since one usually needs to consider
correspondences up to some notion of equivalence relation. This is related to
the fact that the composition of correspondences is given geometrically by
an intersection product: for $\Gamma_1\subset Z \times Y$ and $\Gamma_2 \subset Y \times X$
the composite $\Gamma_2\circ \Gamma_1 \subset Z\times X$ is obtained by pulling
back both correspondences to the triple product $Z\times Y\times X$ via the respective projections,
intersecting them, and pushing the result forward to $Z\times X$,
$$ \Gamma_2\circ \Gamma_1 ={\pi_{Z\times X}}_* (\pi_{Z\times Y}^*(\Gamma_1)\bullet 
\pi_{Y\times X}^*(\Gamma_2)). $$
We do not discuss this in detail here, but some of the subtleties involve having to work with
smooth projective varieties rather than affine schemes, having the freedom to move
representative algebraic cycles describing the correspondences within an equivalence
class that preserves intersection numbers, etc. All these technical aspects are worked
out in the construction of categories of (pure) motives, where morphisms are
described in this way. 
We will not deal directly with motives here, and we simply refer the reader to \cite{Andre}
for a detailed introduction. 

\smallskip

For our purposes, it suffices to point out that the notion of height of a morphism
$\sigma: Y \to X$ that we discussed above, based on measuring the height of a polynomial,
can be generalized to a notion of height of a variety, which can then be applied to
a correspondence $\Gamma$, seen as a subvariety of $Y\times X$. We are making here
a simplifying assumption and considering varieties instead of more general schemes. 
Notions of height function for motives are discussed in \cite{Kato}.

\smallskip

We will not be dealing directly with correspondences in the commutative case,
but we mention this setting as a comparison, because our approach to the
noncommutative case, which we will be discussing shortly, is based on the
analog of correspondences for noncommutative spaces, which is given by
bimodules. 

\medskip

\section{Bimodules and a Noncommutative Functor of Points}\label{BimodPtsSec}

When we move from the commutative to the noncommutative world, the
notion of functor of points as usually constructed in algebraic geometry is
clearly inadequate. If $R$ and $S$ are noncommutative $\K$-algebras,
there are in general very few algebra homomorphisms $\phi: R \to S$,
hence defining a functor of points using $\Hom_{{\rm Alg}_\K}(R,S)$
gives very little information. There are various ways in which noncommutative
geometry has considered the problem of a good definition of points.
A widely used approach consists of replacing the notion of points as
algebra homomorphisms with the notion of points as extremal measures
(delta measures supported on points in the classical commutative case).
In noncommutative geometry, this involves the notion of states on the
algebra of functions. In the case of a complex involutive unital algebra $R_\C$, 
a state is a linear functional $\varphi: R_\C \to \C$ that is normalized $\varphi(1)=1$
and satisfies a positivity condition $\varphi(a^* a)\geq 0$ for all $a\in R_\C$. These
two properties generalize to the noncommutative setting the notion of a measure.
In contrast to the set of algebra homomorphisms, which tends to be too
small in the noncommutative case, the set of extremal points of the convex set of
states tends to be too large, but this problem is usually cured by introducing
additional requirement, for example only considering special states that
are equilibrium (KMS) states for a dynamical evolution of the noncommutative
algebra. This approach was successfully used in applications of
noncommutative geometry to number-theoretic settings, see for instance
Chapters~3 and~4 of \cite{CoMa}. However, this is not the main viewpoint that
we want to consider in this paper. 

\smallskip

We want to highlight here the idea that
the category of sets is not a good category in which to formulate a notion
of points for noncommutative spaces. There are many instances in
modern mathematics where it is clear that categories are a natural
replacement for sets. Thus, we consider the possibility of a functor of
points with values in the category of small categories.

\smallskip
\subsection{Bimodules, 2-categories, and a functor of points}

Again we fix a field $\K$, which we take to be a number field. As a category
of algebras over $\K$ we consider the following.

\begin{defn}\label{NCalgs}
Let $\cN\cA_\K$ denote the category with objects given by associative (not necessarily commutative)
algebras $R$ over $\K$ and morphisms $\Hom_{\cN\cA_\K}(R,S)$ given by bimodules $_{R}E_S$.
The composition of morphisms is given by the tensor product $_{R}E_S \otimes_S {}_S F_{T}=: {}_R F\circ E_T$.
\end{defn}

The associativity of the tensor product of bimodules can be seen in the following way. 
Given $E_1, E_2, E_3$, with $E_i$ an $R_{i-1}-R_i$ module, the tensor product 
$E=E_1\otimes_{R_1} E_2\otimes_{R_2} E_3$ is an $R_0-R_3$ bimodule with a multilinear
map $\mu: E_1 \times E_2 \times E_3 \to E$ from the underlying product of sets that is
$\K$-linear in each variable and satisfies $a_0 \mu(e_1,e_2,e_3)=\mu(a_0 e_1, e_2, e_3)$,
$\mu(e_1,e_2,e_3) a_3=\mu(e_1,e_2,e_3 a_3)$, $\mu(e_1 a_1, e_2, e_3)=\mu(e_1, a_1 e_2,e_3)$,
$\mu(e_1,e_2 a_2, e_3)=\mu(e_1,e_2,a_2 e_3)$, for all $a_i\in R_i$ and $e_i\in E_i$. The
map $\mu$ determines uniquely maps $\mu_1: (E_1\otimes_{R_1} E_2)\times E_3 \to E$
and $\mu_2: E_1 \times (E_2\otimes_{R_2} E_3) \to E$ and is in turn uniquely determined
by each of them. 

\begin{rem}\label{2catNA}{\rm 
Since $R-S$ bimodules form a category, the $\Hom_{\cN\cA_\K}(R,S)$ are categories, hence
$\cN\cA_\K$ is a category enriched over categories, that is, a strict $2$-category. }
\end{rem}

\smallskip

We then propose the following definition of functor of points for noncommutative spaces.
Let ${\rm Cat}$ denote the category of small categories. This is also a $2$-category with
objects that are small categories, morphisms that are functors, and $2$-morphisms that
are natural transformations. 

\begin{lem}\label{NCfunctpoints}
For $R$ an associative (noncommutative) algebra over $\K$, 
let $\pi_R: \cN\cA_\K \to {\rm Cat}$ be the functor that assigns to an object $S$ in $\cN\cA_\K$
the category ${}_R \cB_S$ of $R-S$ bimodules. A morphism ${}_{S_1} F_{S_2} \in \Hom_{\cN\cA_\K}(S_1,S_2)$
is mapped to the functor $F: {}_R \cB_{S_1} \to {}_R \cB_{S_2}$ that sends ${}_R E_{S_1}\mapsto
{}_R E_{S_1}\otimes_{S_1} {}_{S_1}F_{S_2}$.
\end{lem}

\proof Recall that in order for $\pi_R: \cN\cA_\K \to {\rm Cat}$ to be a 2-functor, it must be an assignment of $0$-cells (objects), $1$-cells, and $2$-cells in $\cN\cA_\K$ to those in $\rm Cat$ that strictly preserves identity $1$-cells,
identity $2$-cells, vertical compositions of $2$-cells, and horizontal compositions of
$1$-cells and of $2$-cells \cite{Johnson}.  

\begin{itemize}
    \item For $0$-cells $S$ in $\cN\cA_\K$, $\pi_R$ assigns the category ${}_R \cB_S$ of $R-S$ bimodules;
    \item For $1$-cells ${}_{S_1} F_{S_2} \in \Hom_{\cN\cA_\K}(S_1,S_2)$, $\pi_R$ assigns the functor $F: {}_R \cB_{S_1} \to {}_R \cB_{S_2}$ that sends ${}_R E_{S_1}\mapsto
{}_R E_{S_1}\otimes_{S_1} {}_{S_1}F_{S_2}$;
\item For $2$-cells $f:{}_{S_1} F_{S_2} \to {}_{S_1} F^{'}_{S_2}$ in $\cN\cA_\K$, $\pi_R$ assigns the following natural transformation $\eta: F \to F^{'} $, defined for any ${}_R E_{S_1} \in {}_R \cB_{S_1}$, $e_{i} \in {}_R E_{S_1}$ and $m_{i} \in {}_{S_1} F_{S_2}$
\begin{align*}
    \eta_{{}_R E_{S_1}}\big( \sum_{i} e_{i}\otimes m_{i}\big)= \sum_{i} e_{i}\otimes f(m_{i})
\end{align*}
This is a natural transformation since for any bimodule homomorphism $g: {}_R E_{S_1}\to {}_R E^{'}_{S_1}$, 
\begin{align*}
    F^{'}(g)\circ \eta_{{}_R E_{S_1}}\big( \sum_{i} e_{i}\otimes m_{i}\big)&=F^{'}(g)\big(\sum_{i} e_{i}\otimes f(m_{i})\big)\\
    &= \sum_{i} g(e_{i})\otimes f(m_{i})\\
    &=\eta_{{}_R E^{'}_{S_1}}\big(\sum_{i} g(e_{i})\otimes m_{i}\big)\\
    &= \eta_{{}_R E^{'}_{S_1}} \circ F(g) \big(\sum_{i} e_{i}\otimes m_{i}\big)
\end{align*}
\item $\pi_R$ preserves identity $1$-cells since for any $0$-cell $S$ in $\cN\cA_\K$, $\pi_R$ takes its identity $1$-cell ${}_{S}S_{S}$  to the functor $S: {}_R \cB_{S} \to {}_R \cB_{S}$ that sends ${}_R E_{S}\mapsto
{}_R E_{S}\otimes_{S} {}_{S}S_{S}$. Since ${}_R E_{S}\otimes_{S} {}_{S}S_{S}= {}_R E_{S}$, the functor $S: {}_R \cB_{S} \to {}_R \cB_{S}$ is the identity functor of ${}_R \cB_{S}$;
\item $\pi_R$ preserves identity $2$-cells since any identity bimodule homomorphism $f:{}_{S_1} F_{S_2} \to {}_{S_1} F_{S_2}$ is sent to the natural transformation $\eta: F \to F $, defined for any ${}_R E_{S_1} \in {}_R \cB_{S_1}$, $e_{i} \in {}_R E_{S_1}$ and $m_{i} \in {}_{S_1} F_{S_2}$
\begin{align*}
    \eta_{{}_R E_{S_1}}\big( \sum_{i} e_{i}\otimes m_{i}\big)&= \sum_{i} e_{i}\otimes f(m_{i})\\
    &= \sum_{i} e_{i}\otimes m_{i}
\end{align*}
which is precisely an identity natural transformation since it maps each object ${}_R E_{S_1}$ of $ {}_R \cB_{S_1}$ to the identity morphism $\text{id}_{F({}_R E_{S_1})}$ in ${}_R \cB_{S_2}$; 
\item For any two composable bimodules ${}_{S_1} F_{S_2}$,  ${}_{S_2} G_{S_3}$, 
\begin{align*}
  \pi_R({}_{S_{1}} F\circ G_{S_{3}})&=  \pi_R({}_{S_{1}} F_{S_2} \otimes_{S_{2}} {}_{S_{2}} G_{S_{3}})\\
  &=- \otimes_{S_1} {}_{S_{1}} F_{S_2} \otimes_{S_{2}} {}_{S_{2}} G_{S_{3}}\\
  &= (- \otimes_{S_1} {}_{S_{1}} F_{S_2}) \circ  (-\otimes_{S_{2}} {}_{S_{2}} G_{S_{3}})\\
  &= \pi_R({}_{S_{1}} F_{S_2}) \circ \pi_R({}_{S_{2}} G_{S_{3}})
\end{align*}
showing that $\pi_R$ preserves the horizontal compositions of 1-cells;
\item  For two vertically composable bimodule homomorphisms $f: {}_{S_{1}} F_{S_2} \to  {}_{S_{1}} G_{S_2}$ and $g: {}_{S_{1}} G_{S_2} \to {}_{S_{1}} H_{S_2}$ in $\cN\cA_\K$, and ${}_R E_{S_1} \in {}_R \cB_{S_1}$, $e_{i} \in {}_R E_{S_1}$ and $m_{i} \in {}_{S_1} F_{S_2}$
\begin{align*}
   \pi_R(g \circ f) _{{}_R E_{S_1}}\big( \sum_{i} e_{i}\otimes m_{i}\big)&= \sum_{i} e_{i}\otimes (g \circ f)(m_{i})\\
    &= \pi_R(g) _{{}_R E_{S_1}}\big(\sum_{i} e_{i}\otimes f(m_{i})\big)\\
    &= (\pi_R(g) \circ \pi_R(f))_{{}_R E_{S_1}}\big(\sum_{i} e_{i}\otimes m_{i}\big)
\end{align*}
\item Finally, for two horizontally composable bimodule homomorphisms $f: {}_{S_1} F_{S_2} \to {}_{S_1} F^{'}_{S_2}$
and $g : {}_{S_2} G_{S_3} \to {}_{S_2} G^{'}_{S_3}$, and  ${}_R E_{S_1} \in {}_R \cB_{S_1}$, $e_{i} \in {}_R E_{S_1}$, $m_{i} \in {}_{S_1} F_{S_2}$, $n_{i} \in {}_{S_2} G_{S_3}$, 
\begin{align*}
   \pi_R(f \otimes g) _{{}_R E_{S_1}}\big( \sum_{i} e_{i}\otimes m_{i}\otimes n_{i}\big)&=\sum_{i} e_{i}\otimes f(m_{i})\otimes g(n_{i})\\
    &= \pi_R(f) _{{}_R E_{S_1}}\big(\sum_{i} e_{i}\otimes m_{i} \otimes g(n_{i}) \big)\\
    &= (\pi_R(f) \circ \pi_R(g))_{{}_R E_{S_1}}\big(\sum_{i} e_{i}\otimes m_{i}\otimes n_{i}\big)
\end{align*}

\end{itemize}
\endproof 

The previous result motivates the following definition.

\begin{defn}\label{defNCfunctpoints}
Let $X_R^{nc}$ denote the noncommutative space determined by the algebra $R$ over $\K$.
We define the noncommutative functor of points by setting $X_R^{nc}(S):={}_R \cB_S$
for any $\K$-algebra $S$.
\end{defn}

Thus, the $S$-points $X_R^{nc}(S)$ of the noncommutative space $X_R^{nc}$ form a $2$-category rather than
a set and consist of all $R-S$ bimodules. 

\smallskip

Thus, we consider $X_R^{nc}(S)={}_R \cB_S$ to be our kinematic space for noncommutative
sigma models, where the noncommutative space $X_R^{nc}$ is the (fixed) target space of the sigma
model and the noncommutative space determined by the algebra $S$ is the (variable) source space
of the sigma model.

\smallskip

The next step is then to make these spaces dynamical by assigning an action functional that
generalizes the height action functional we have been discussing in the previous section. 

\medskip

\section{The case of finite dimensional algebras}\label{FinDimSec}

Before discussing the general problem of how to obtain an action functional on 
on our noncommutative ``functor of points" $X_R^{nc}(S)={}_R \cB_S$, we
focus on the simpler case of finite dimensional algebras. 

\smallskip

It is convenient to adopt the viewpoint of Arakelov geometry in a noncommutative setting, \cite{Borek1}, \cite{Borek2}. 
Consider a finite dimensional semisimple algebras $A$ over a number field $\bK$. 
In the case where $A$ is a division algebra, the construction of a height function
for free submodules of $A^n$ was obtained in \cite{LiRe}, generalizing the
commutative construction of heights of subspaces of a vector space over a
number field using volumes of Euclidean lattices (see \cite{Schm}).
If $A$ is not a division algebra, then it has zero-divisors, and this prevents the
usual construction of valuations. This case is analyzed in \cite{Borek1}, where
a different way of defining a height function $h_\cO(V)$ of a free submodule of $A^n$
is introduced, which for division algebras agrees up to a scale factor with the height 
defined in \cite{LiRe}.

\smallskip

The general structure of finite dimensional algebras is analyzed in \cite{Drozd}.
For our purposes, we consider the case of semi-simple algebras, since  
these algebras can be viewed as sums of simple algebras, and this
naturally generalizes to the number field case the much simpler complex
case where one deals with sums of matrix algebras. Indeed, every simple
algebra is isomorphic to a matrix algebra $M_n(D)$ over some division
algebra $D$ over $\K$.

\smallskip

We show here how to adapt in our setting this approach to associate an
action functional defined on our $X_R^{nc}(S)={}_R \cB_S$, in the case
where both $R$ and $S$ are semisimple finite dimensional algebras 
over a number field $\K$. 

\smallskip
\subsection{Noncommutative arithmetic curves}

In arithmetic geometry, an arithmetic (affine) curve is ${\rm Spec}(\cO_\K)$,
for a number field $\K$ with ring of integers $\cO_\K$. The curve is ``compactified"
by adding to ${\rm Spec}(\cO_\K)$ the ``primes at infinity", that is, the $n$
embeddings $\sigma: \K \hookrightarrow \C$ for $n=[\K:\Q]=\deg(\K)$ the degree of the
number field. Heuristically, one thinks of an arithmetic curve ${\rm Spec}(\cO_\K)$
as having at each finite place $\wp$ in ${\rm Spec}(\cO_\K)$ a copy of the
corresponding residue field $\F_q =\cO_\K/\wp$. (At the archimedean places
these should be replaced by a more mysterious object $\F_1$, which has many
different mathematical incarnations, \cite{LoLo}, \cite{Man2}.)

\smallskip

In noncommutative algebraic geometry there is a similar notion
of an arithmetic noncommutative curve, see \cite{Borek1}, which we
review briefly here.

\begin{defn}\label{orderOA}{\rm
Given a semisimple finite dimensional algebra $A$ over a number field $\K$,
an $\cO_\K$-order $\cO_A$ in $A$ is a subring of $A$ such that $\cO_A$
is a full $\cO_\K$-lattice in $A$. 
}\end{defn}

This means that $\cO_A$ is a finitely generated 
torsion free $\cO_\K$-module with the property that $\K \cO_A=A$ as $\K$-vector spaces.

\smallskip

A notion of arithmetic noncommutative curve is then provided by 
${\rm Spec}(\cO_A)$ where $\cO_A$ is an $\cO_\K$-order in $A$.
The reason why it makes sense to consider this a noncommutative
arithmetic curve lies in the fact that the prime ideals $\wp$ in ${\rm Spec}(\cO_A)$
are maximal two-sided ideas of $\cO_A$ and the quotients $\cO_A/\wp$ are simple
algebras hence isomorphic to some matrix algebra $M_{k_\wp}(D_\wp)$ over some
division algebra $D_\wp$. Thus, this construction replaces the commutative objects
$\F_q$ attached to the places $\wp$ of ${\rm Spec}(\cO_\K)$ in a classical (commutative)
arithmetic curve, with the noncommutative objects $M_{k_\wp}(D_\wp)$ attached
to the places $\wp$ of the noncommutative curve ${\rm Spec}(\cO_A)$. 

\smallskip

While this point of view is very helpful, it is slightly different from the one we
will be following here, since it is still based on taking two-sided ideals $\wp$,
related to a classical notion of points. We will however, keep this main 
idea in mind in framing our setting, while adapting it to our notion of points
described in the previous section. 

\smallskip

Before developing our setting, we recall a few more aspects of the
theory of noncommutative arithmetic curves, as developed by Borek in \cite{Borek1},
which will be useful in our setting as well, in particular some notions of height
that we will be adapting to our setting in the next subsections.

\smallskip

For a semisimple finite dimensional algebra $A$ over a number field $\K$,
We write $A_\R:=A\otimes_\K \R$ and $A^*_\R:=\GL_1(A_\R)$ for the group
of units (multiplicatively invertible elements) in $A_\R$.  For a $\cO_\K$-order 
$\cO_A$ in $A$, we also write $\cJ(\cO_A)$ for the set of full left-$\cO_A$-ideals in $A$. 
These are left $\cO_A$-modules that are full $\cO_\K$-lattices.  For a full left-
$\cO_A$-ideal $\fa$ in $A$ there is an $r\in \cO_\K$ such that
$\fa r \subset \cO_A$ is a left-ideal in $\cO_A$.  The set of complete
$\cO_A$-ideals is $\hat\cJ(\cO_A)=\cJ(\cO_A)\times A_\R^*$. We write complete
$\cO_A$-ideals as $\bar\fa=(\fa,\fa_\infty)\in \cJ(\cO_A)\times A_\R^*$, as in \cite{Borek1}.
For $\fa\in \cJ(\cO_A)$ one takes 
\begin{equation}\label{Na}
\cN(\fa) = \frac{\# (\cO_A/\fa r)}{\# (\cO_A/\cO_A\, r)}. 
\end{equation} 
This is independent of the choice of an $r\in \cO_\K$ for which $\fa r$ is a left-ideal in $\cO_A$. 
As shown in Theorem~1 of \cite{Borek1}, the norm $\cN(\fa)$ of \eqref{Na}
has an equivalent expression as product over the prime ideals $\wp$ in ${\rm Spec}(\cO_A)$. 
The ``absolute norm" $\cN(\bar \fa)$ is given by the product of $\cN(\fa)$ and a contribution
of the archimedean component $\fa_\infty$ given by $|\, N_{A_\R|\R}(\fa_\infty)\,|$ where
$N_{A_\R|\R}$ is the norm map from $A_\R$ to $\R$. (In general, given a finite dimensional
algebra $B$ over a field $F$ an element $x\in B$ acts by left multiplication defining
an $F$-endomorphism of $B$. The norm $N_{B|F}: B \to F$ is the multiplicative map given by 
$N_{B|F}(x)=\det(x)$, the determinant of
the resulting $F$-linear transformation.) 

\smallskip

One can also define a volume ${\rm vol}(\bar \fa):=
{\rm vol}(\fa_\infty j(\fa))$, where $j:A \to A_\R$ maps $\fa$ to a $\Z$-lattice $j(\fa)$ in $A_\R$
with $\R j(\fa)=A_\R$, and the volume of the lattice $\fa_\infty j(\fa)$ is the volume of its
fundamental domain in the measure defined by the norm associated to the bilinear form
$A_\R\times A_\R \to \R$ determined by the trace $(x,y)\mapsto \Tr_{A_\R|\R}(xy)$.
This is related to the absolute norm by ${\rm vol}(\bar \fa)=\cN(\bar\fa)\, {\rm vol}(\cO_A)$,
see \S 4 of \cite{Borek1}. We will discuss more generally in \S \ref{VolSec1} and \S \ref{VolSec} the
assignment of a volume to a normed space and notions of volume we will be using in the
construction of height functions. 

\smallskip

The last notions that we need to recall from the Arakelov geometry of noncommutative
curves of \cite{Borek1} is arithmetic vector bundles and the associated arithmetic degree map.
Again, the notion of arithmetic vector bundle that we will be using is slightly different from
the one used in \cite{Borek1} that we recall here, but it is closely related.

\smallskip

\begin{defn}\label{hdefin} {\rm 
Let $\K$ be a number field and $A$ a semisimple finite dimensional algebra over $\K$,
with $\cO_A$ an order in $A$. Let $\cE$ be a left $\cO_A$-module. Let $\sigma$ be an archimedean
place of $\K$ and let $\K_\sigma$ be either $\R$ or $\C$, with the corresponding embedding
$\sigma: \K \hookrightarrow \K_\sigma$, and let $A_\sigma=\K_\sigma\otimes_\K A$ and
$\cE_\sigma=A_\sigma\otimes_{\cO_A} \cE$. 
A hermitian structure $h$ on $\cE$ is a $\star$-hermitian
bilinear form $h: \cE_\sigma \times \cE_\sigma \to A_\sigma$ such that
$\Tr_{A_\sigma|\K_\sigma}\circ h: \cE_\sigma\times \cE_\sigma \to \K_\sigma$, 
$(x,y)\mapsto \Tr_{A_\sigma|\K_\sigma}(h(x,y))$ is positive definite. 
}\end{defn}

\smallskip

Such hermitian metrics can be constructed in the following way, \cite{Borek1}.
The verification of the following statement is immediate and we omit it.

\begin{lem}\label{hbeta}
The choice of a collection of elements $\beta_i \in A_\sigma^*$, for $i=1,\ldots, n$ 
determines a hermitian structure on the free module $A_\sigma^n$ by setting 
$h(x,y)=\sum_i x_i \beta_i \beta_i ^*y_i^*$. By restriction, it determines a hermitian
structure on a projective $A_\sigma$-module that is a summand of the free module $A_\sigma^n$.
\end{lem}

In more physical terms, we can regard $\rho_i=\beta_i \beta_i^*$ as a ``density matrix" 
and $\Tr_{A_\sigma|\K_\sigma}\circ h (x,y)=\Tr(y^* x \rho)$ as the state associated to this density matrix.

\smallskip

\begin{defn}\label{standardh}{\rm 
Let $\cE$ be a finite projective left $A$-module. For $\sigma$ an archimedean place of $\K$,
a hermitian structure $h: \cE_\sigma \times \cE_\sigma \to \K_\sigma$ is said to be ``standard"
if it is the restriction of a hermitian form as in Lemma~\ref{hbeta} on $A_\sigma^n$, for an embedding
$j_A: \cE_\sigma \hookrightarrow A_\sigma^n$ realizing $\cE_\sigma$ as a summand of a free module.}
\end{defn}

\smallskip

It is shown in \cite{Borek1} that all hermitian structures $h: A_\R \times A_\R \to A_\R$ 
on a real semisimple finite dimensional algebra are of the standard 
form $h: (x,y)\mapsto x \beta \beta^* y^*$ as above. The standard hermitian structures
will be sufficient for our purposes, so we will limit some of the arguments to this case.

\smallskip

An arithmetic vector bundle over the noncommutative arithmetic curve ${\rm Spec}(\cO_A)$
is defined in \cite{Borek1} as a pair $(\cE,h)$, where $\cE$ is a left $\cO_A$-module that is also an $\cO_\K$-lattice,
such that $A\otimes_{\cO_A}\cE$ is a free $A$-module, 
and $h$ is a hermitian structure $h: \cE_\R \times \cE_\R \to A_\R$
on $\cE_\R=A_\R \otimes_{\cO_A} \cE$. Our notion of arithmetic vector bundle is
discussed in Definition~\ref{arVB} below.

\smallskip

The Grothendieck group $K_0({\rm Spec}(\cO_A))$ of arithmetic vector bundles of \cite{Borek1}
over the noncommutative arithmetic curve ${\rm Spec}(\cO_A)$ is generated by
isomorphism classes of $(\cE,h)$, where isomorphism is given by isomorphisms
of left $\cO_A$-modules that preserve the hermitian structure, modulo the relations
$[(\cE,h)]=[(\cE',h')]+[(\cE'',h'')]$ for short exact sequences  
$0\to \cE'\to \cE \to \cE'' \to 0$ of $\cO_A$-modules
such that the sequence
$0\to \cE_\R'\to \cE_\R \to \cE_\R'' \to 0$
splits orthogonally, that is, with $\cE_\R'' =(\cE'_\R)^\perp$ the orthogonal
with respect to the hermitian structure. 

\smallskip

Complete $\cO_A$-ideals $\bar\fa\in \hat\cJ(\cO_A)$ determine
hermitian line bundles $\cL(\bar\fa)$ that generate the Grothendieck 
group $K_0({\rm Spec}(\cO_A))$. The degree map
$$ \deg_{\cO_A}(\cL(\bar\fa)):= -\log \cN(\bar\fa) $$
extends uniquely to a degree map on $K_0({\rm Spec}(\cO_A))$. 

\smallskip

Finally, we recall the notion of height used in the Arakelov geometry of
noncommutative curves, \cite{Borek1}. Let $h_n$ denote the hermitian
metric on $A^n_\R$ given by $h_n: ((x_1,\ldots,x_N),(y_1,\ldots,y_N))\mapsto \sum_i x_i^* y_i$.
For a free $A$-submodule $\cV$ of $A^n$ the logarithmic height is defined as
$$ h_{\cO_A}(\cV)= \log H_{\cO_A}(\cV) := - \deg_{\cO_A} (\cV\cap \cO_A^n, h_n). $$
By the relation of the norm to the volume, the height can be further identified with
\begin{equation}\label{BorekH}
  H_{\cO_A}(\cV) = \frac{{\rm vol}(\cV\cap \cO_A^n)}{{\rm vol}(\cO_A)^{{\rm rank}(\cV)}}, 
\end{equation} 
where the volume in $A^n_\R$ is defined by the norm associated to the quadratic 
form $\Tr_{A_\R|\R}\circ h_n$. This generalizes to the noncommutative setting the
height of \cite{Schm}. In the case of division algebras it agrees (up to scaling factor
given by the degree $[\K:\Q]$) with the height of \cite{Lieb}, \cite{LiRe}, \cite{LiRe2}.
Here the hypotheses that an arithmetic vector bundle is a free $A$-module as well as
an $\cO_\K$-lattice, assumed in \cite{Borek1} are used in defining the volume 
of $\cV\cap \cO_A^n$ and the rank of $\cV$. We will discuss how this definition
of height adapts to our setting in the next subsections. 

\smallskip
\subsection{Free modules versus finite projective modules}

The notion of arithmetic vector bundle over a noncommutative arithmetic curve,
that we recalled above from \cite{Borek1}, includes the requirement that the
left $\cO_A$-module $\cE$ satisfies the property that $A\otimes_{\cO_A} \cE$ is
a free $A$-module. 

\smallskip

It would seem more natural, if we want to regard $\cE$ as a vector bundle, to
relax this requirement and only require that $A\otimes_{\cO_A} \cE$ is a finite
projective $A$-module. As we see in the following, in our setting based on
bimodules, this less restrictive condition will allow us to include some very
natural families of bimodules that we certainly want to include in our counting
of ``points" over our noncommutative arithmetic spaces. 

\smallskip

Thus, in our setting, we give the following slightly different definition of
an arithmetic vector bundle. 

\begin{defn}\label{arVB}
Let $A$ be a semisimple finite dimensional algebra over a number field $\K$ and
let $\cO_A$ be an order in $A$. A left arithmetic $(A,\cO_A)$-vector bundle $(\cE,h)$ 
is a left $\cO_A$-module, together with the assignment $h=(h_\sigma)$, for all archimedean places
$\sigma$ of $\K$, of a hermitian structure $h: \cE_\sigma \times \cE_\sigma \to A_\sigma$.
\end{defn}

Since $A$ is a semisimple finite dimensional algebra, the $A$-module $A\otimes_{\cO_\K} \cE$
is finite projective (see also Remark~\ref{finprojcond} below).

\smallskip
\subsection{Functor of points on finite dimensional algebras}

We now adapt the notions developed in the Arakelov geometry of noncommutative
arithmetic curves in \cite{Borek1} to our setting.

\smallskip

Let $A$ be a semisimple finite dimensional algebra over a number field $\K$.
Let $\sigma: \K \hookrightarrow \C$ be an embedding (either real or complex),
that is, an archimedean place of $\K$. We can associate to $A$ and $\sigma$
a real or complex algebra $A_\sigma =A\otimes_{\K,\sigma} \R$ for a real embedding
and $A_\sigma =A\otimes_{\K,\sigma}\C$ for a complex embedding. Assuming
$A=\oplus_i M_{n_i}(D_i)$ where the simple summands are matrix algebras of
rank $n_i$ over a division algebra $D_i$ over $\K$, we can similarly decompose
$A_\sigma=\oplus_i M_{m_i}(\C)$ in the complex case and $A_\sigma=\oplus_i M_{r_i}(\R)
\oplus_j M_{r_j}(\C) \oplus_k M_{r_k}(\H)$ for the real case, where $\H$ is the
division algebra of quaternions. 

\smallskip

\begin{defn}\label{Hbimods} {\rm
An {\em arithmetic structure} on a semisimple finite dimensional algebra $A$
over a number field $\K$ is the choice of an $\cO_\K$-order $\cO_A$ in $A$.
Given two such a choices $(A,\cO_A)$ and $(B,\cO_B)$, we define the
category of arithmetic hermitian bimodules $_{(A,\cO_A)}\cH_{(B,\cO_B)}$ with
objects the pairs $(\cE,h)$ where $\cE$ is an $\cO_A-\cO_B$ bimodule,
such that $A\otimes_{\cO_A} \cE$ and $\cE\otimes_{\cO_B} B$  are finite projective
as $A$-module and $B$-module, respectively. 
Moreover, for any archimedean place $\sigma$ of $\K$ the
bimodule $\cE_\sigma:=A_\sigma \otimes_{\cO_A} 
\cE \otimes_{\cO_B} B_\sigma$ is endowed with a pair of hermitian 
structures $h=(h_A,h_B)$ with $h_A: \cE_\sigma \times \cE_\sigma \to A_\sigma$
and $h_B: \cE_\sigma \times \cE_\sigma \to B_\sigma$ with the following properties
\begin{itemize}
\item For all $a\in A_\sigma$, the identities 
$h_A(a x,y)=a \, h_A(x,y)$ and $h_A(x,ay)=h_A(x,y)\, a^*$ hold for
all $(x,y)\in \cE_\sigma \times \cE_\sigma$.
\item For all $x\in \cE_\sigma$ the element $h_A(x,x)$ is a positive element
in $A_\sigma$ (that is, an element of the form $\beta^* \beta$ for some $\beta\in A_\sigma$
and $h_A(x,x)=0$ iff $x=0$.  
\item For all $b \in B_\sigma$, the identities $h_B(x,yb)=h_B(x,y)\, b$ and $h_B(xb,y)=b^*\, h_B(x,y)$
hold for all $(x,y)\in \cE_\sigma \times \cE_\sigma$.
\item For all $x\in \cE_\sigma$ the element $h_B(x,x)$ is a positive element
in $B_\sigma$ and $h_B(x,x)=0$ iff $x=0$. 
\end{itemize}
Morphisms $\phi: (\cE,h_A,h_B)\to (\cE', h_A', h_B')$ are morphisms $\phi:\cE\to \cE'$ 
of $\cO_A-\cO_B$ bimodules such that the induced morphisms $\phi_\sigma: \cE_\sigma\to \cE'_\sigma$
satisfy $h_A'(\phi_\sigma(x),\phi_\sigma(y))=h_A(x,y)$ and $h_B'(\phi_\sigma(x),\phi_\sigma(y))=h_B(x,y)$.
}\end{defn}

\smallskip

\begin{defn}\label{Hbimods2} {\rm
Consider an arithmetic structure 
on a semisimple finite dimensional algebra $A$
over a number field $\K$ as in Definition~\ref{Hbimods}. A {\em strong} arithmetic hermitian bimodule
$(\cE,h)$ is an object of $_{(A,\cO_A)}\cH_{(B,\cO_B)}$ as in Definition~\ref{Hbimods} with
the additional properties that $\cE$ is also left-right $\cO_\K$-lattice, such that $A\otimes_{\cO_A} \cE$ is
a free $A$-module and $\cE\otimes_{\cO_B} B$ is a free $B$-module.
}\end{defn}

\smallskip

Definition~\ref{Hbimods2} more closely matches the notion of arithmetic
vector bundles used in \cite{Borek1}. We prefer here to consider the more
general class of bimodules of Definition~\ref{Hbimods}.  

\begin{rem}\label{finprojcond} {\rm
The condition that $A\otimes_{\cO_A} \cE$ and $\cE\otimes_{\cO_B} B$  
are finite projective modules in Definition~\ref{Hbimods} is automatically
satisfied since $A$ and $B$ are semisimple algebras (\cite{Lam}, Theorem~2.8 
and Corollary~3.7).
}\end{rem}

\smallskip

Given such a choice of a pair $(A,\cO_A)$, and another $(B,\cO_B)$,
we define as in the previous section the noncommutative functor of
points $X^{nc}_{(A,\cO_A)}(B,\cO_B)$ (the  $(B,\cO_B)$-``points" of the
noncommutative space defined by $(A,\cO_A)$) to be the category
of hermitian bimodules defined as above, 
$$X^{nc}_{(A,\cO_A)}(B,\cO_B) := \,\,  {}_{(A,\cO_A)}\cH_{(B,\cO_B)}. $$ 

\smallskip

We refer here to arithmetic vector bundles over ${\rm Spec}(\cO_A)$
as {\em left arithmetic vector bundles} and we similarly define {\em right arithmetic
vector bundles} $(\cE,h)$ in the same way but with $\cE$ a right $\cO_A$-module.

\smallskip
\subsection{Hermitian structures on bimodules}

We discuss here some explicit construction of hermitian structures on bimodules
between semisimple finite dimensional algebras, which will be useful later in the
construction of the height function.

\smallskip

\begin{lem}\label{Hbilem}
As above, let $(A,\cO_A)$ and $(B,\cO_B)$ be pairs of semisimple finite
dimensional algebras over a number field $\K$ and orders. Let $(\cE_A,h^{\cE_A}_A)$
be a left arithmetic $(A,\cO_A)$-vector bundle and let $(\cE_B,h^{\cE_B}_B)$ be a right arithmetic 
$(B,\cO_B)$-vector bundle, where the hermitian structures are standard as in Definition~\ref{standardh}.
Then $\cE=\cE_A\otimes_\K \cE_B$ is an arithmetic hermitian bimodule with standard hermitian structures.
\end{lem}

\proof  
$\cE=\cE_A\otimes_\K \cE_B$
is a hermitian bimodule if it satisfies all the conditions in Definition \ref{Hbimods}. Given that $(\cE_A,h_A)$ and $(\cE_B,h_B)$ are left and right arithmetic vector bundles, respectively, then $\cE=\cE_A\otimes_\K \cE_B$ is an $\cO_A-\cO_B$ bimodule. As noted in Remark~\ref{finprojcond}, the condition
that $A\otimes_{\cO_A} \cE$ and $\cE\otimes_{\cO_B} B$  
are finite projective modules is automatically satisfied.

For any archimedean place $\sigma$ of $\K$, one needs to show that the
bimodule $\cE_\sigma:=A_\sigma \otimes_{\cO_A} 
\cE \otimes_{\cO_B} B_\sigma$ can be endowed with a pair of hermitian 
structures $h=(h_A,h_B)$ with $h_A: \cE_\sigma \times \cE_\sigma \to A_\sigma$
and $h_B: \cE_\sigma \times \cE_\sigma \to B_\sigma$ satisfying all the conditions in Definition \ref{Hbimods}. 

Consider the given hermitian structure $h^{\cE_A}_A: \cE_{A,\sigma} \times \cE_{A,\sigma} \to A_\sigma$.
We are assuming it is standard in the sense of Definition~\ref{standardh}, hence we have:
\begin{itemize}
\item an integer $n=n_A\in \N$ such that $\cE_A = A^n e_A$ for an idempotent $e_A\in M_n(A)$
\item elements $\beta_{A,i} \in A^*$, for $i=1,\ldots, n$ such that $h^{\cE_A}_A(x,y)=\sum_i x_i \beta_{A,i} \beta_{A,i}^* y_i=: {}_{A}\langle x, y \rangle$
\item elements $\{ u_i \}$ in $\cE_A$ such that $x=\sum_i x_i u_i$ and $y=\sum_i y_i u_i$ with 
$x_i={}_{A}\langle x, u_i \rangle$ and $y_i ={}_{A}\langle y, u_i \rangle$
and with  $(e_A)_{ij}={}_{A}\langle u_i, u_j \rangle$.
\end{itemize}
Consider $\cE_B$ as a vector space over $\K$, and for each $i=1,\ldots, n$ let $\{v_{i,\ell} \}_{\ell \in \cJ}$ 
be a choice of a basis for $\cE_B$ as $\K$-vector space. Then the set $\{ u_i \otimes v_{i,\ell} \}$ of
elements of $\cE_A\otimes_\K \cE_B$ has the property that $\xi =\sum_{i,\ell} x_{i,\ell} \, u_i \otimes v_{i,\ell}$,
for $x\in \cE_A\otimes_\K \cE_B$, with the $x_{i,\ell}\in A$. 

One can then define $h_A$ on $\cE_\sigma$ as follows. We write an element $x\in  \cE_\sigma$ 
as above, with the $x_{i,\ell} \in A_\sigma$ and similarly for $y=\sum_{i,\ell} y_{i,\ell} \, u_i \otimes v_{i,\ell}$.
With a slight abuse of notation we will still denote by $\beta_{A,i}$ as above the diagonal matrix
in $M_r(A)$ with $r=\# J$ with $\beta_{A,i}\in A^*$ on the diagonal. We can then define
$$ h_A(x,y):= \sum_i x_i \beta_{A,i} \beta_{A,i}^* y_i^* =\sum_\ell h_A^{\cE_A}(x_\ell , y_\ell), $$
where $x_i \beta_{A,i} \beta_{A,i}^* y_i^* =\sum_{\ell,\ell'\in \cJ} x_{i,\ell}
 \beta_{A,i,\ell,\ell'} \beta_{A,i,\ell,\ell'}^* y_{i,\ell'}^*$ and $x_\ell :=\sum_i x_{i,\ell} u_i \otimes v_{i,\ell}$,
 $y_\ell :=\sum_i y_{i,\ell} u_i \otimes v_{i,\ell}$. The resulting $h_A$ constructed in this way is
 manifestly also a standard hermitian structure.
 
The construction of the hermitian structure $h_B$ is similarly obtained, using 
$h_{B}^{\cE_B}$, the hermitian metrics coming from the arithmetic vector bundle $\cE_B$. 
One defines $h_B: \cE_\sigma \to B_\sigma$ by setting 
$$ h_B(x,y):= \sum_j x_j \beta_{B,j} \beta_{B,j}^* y_j^* =\sum_\ell h_B^{\cE_B}(x_\ell , y_\ell), $$
where $\beta_{B,j}\in B^*$ are the elements of the standard hermitian structure $h_{B}^{\cE_B}$,
$x=\sum_{\ell \in \cI} v'_{j,\ell}\otimes u'_j\,\,  x_{j,\ell} $ with $\{ v'_{j,\ell} \}_{\ell\in \cI}$ a basis
of $\cE_A$ as a $\K$-vector space and $x_j=(x_{j,\ell})_\ell$ for fixed $j$ and $x_\ell =\sum_j 
v'_{j,\ell}\otimes u'_j\,\,  x_{j,\ell}$ for fixed $\ell$.
Due to the fact that $h_A^{\cE_A}$ and $h_B^{\cE_B}$ are standard hermitian metrics, 
$h_A$ and $h_B$  also satisfy the conditions in Definition \ref{Hbimods} and Definition~\ref{standardh}.. 
\endproof

\smallskip
\subsection{A height function on hermitian bimodules}

We now discuss how to introduce a height function on the hermitian bimodules
in ${}_{(A,\cO_A)}\cH_{(B,\cO_B)}$, which we regard as our action functional
for weighting the bimodules, thought of as ``maps" of a noncommutative
sigma model with target a noncommutative space $X^{nc}_{(A,\cO_A)}$
(which we can think of as another manifestation of the noncommutative 
arithmetic curve ${\rm Spec}(\cO_A)$). We present here the general form the
height function should take and we complete our definition of height 
after a further discussion of norms and volumes in \S \ref{VolSec1}
and \ref{VolSec}.

\smallskip

First observe that, 
in the case of a ``strong" hermitian bimodule, in the sense of Definition~\ref{Hbimods2},
one can proceed as in \cite{Borek1} to define a height function, as we
recalled above, but taking into account the presence of two hermitian
structures $h=(h_A,h_B)$. 

\smallskip

For $\sigma$ an archimedean place of $\K$, let $\K_\sigma:=\R$ if
$\sigma$ is a real embedding and $\K_\sigma:=\C$ if it is a complex
embedding. 

\smallskip

Let ${\rm vol}_{h_A,\sigma}$ denote
a volume in $A^n_\sigma$ determined by the
norm associated to the bilinear form 
$\Tr_{A_\sigma|\K_\sigma}\circ h_A$ (see \S \ref{VolSec1} below for
a discussion of volumes associated to norms.)
Let $\cV$ be a free $A$-submodule of $A^n$ of rank $r_A$.
We write the normalized and non-normalized heights, respectively, as
\begin{equation}\label{HhA}
\bar H_{\cO_A,h_A,\sigma}(\cV):= \frac{{\rm vol}_{h_A,\sigma}(\cV\cap \cO_A^n)}{{\rm vol}_{h_A,\sigma}(\cO_A)^{r_{A,\sigma}}}
\ \ \ \text{ and } H_{\cO_A,h_A,\sigma}(\cV):= {\rm vol}_{h_A,\sigma}(\cV\cap \cO_A^n).
\end{equation}
The normalized height agrees with the height $H_{\cO_A}$ of \cite{Borek1} when $h_A$ is the
standard hermitian metric $h_n$ and $\sigma$ is a real embedding. 

\begin{defn}\label{HEhABdef}{\rm 
Let $(\cE,h_A,h_B)$ be a hermitian bimodule
in ${}_{(A,\cO_A)}\cH_{(B,\cO_B)}$, as in Definition~\ref{Hbimods}. 
We define the normalized and non-normalized heights of $(\cE,h_A,h_B)$ as
\begin{equation}\label{HEhAB}
\bar H(\cE,h_A,h_B):=\prod_{\sigma} \frac{\bar H_{\cO_B,h_B,\sigma}(\cE_B)}{\bar H_{\cO_A,h_A,\sigma}(\cE_A)}
\ \ \ \text{ and } \ \ \  H(\cE,h_A,h_B):=\prod_{\sigma} \frac{H_{\cO_B,h_B,\sigma}(\cE_B)}{H_{\cO_A,h_A,\sigma}(\cE_A)}, 
\end{equation}
where $\sigma$ ranges over the archimedean places of $\K$, with 
$\cE_A=A\otimes_{\cO_A} \cE$ and $\cE_B=\cE\otimes_{\cO_B} B$. 
}\end{defn}

\smallskip

In order to have a meaningful minimization problem for \eqref{HEhAB}
one needs to minimize over classes of hermitian bimodules with a
fixed positive lower bound on $\bar H_{\cO_A,h_A,\sigma}(\cE_A)$ while
minimizing over $\bar H_{\cO_B,h_B,\sigma}$. For modules of fixed rank,
minimizing with respect to the normalized or non-normalized height is the same.

\smallskip

In our setting we also want to consider a notion of relative height in the case 
of finite projective modules instead of free modules, so that Definition~\ref{HEhABdef} above
can be applied to the more general class of hermitian bimodules of Definition~\ref{Hbimods}
and not only to the strong ones. 

\smallskip
\subsection{The rank element}\label{rankSec}

We will discuss more in detail how to obtain a suitable 
volume computation in \S \ref{VolSec1} and \S \ref{VolSec}.
We discuss here the rank $r_A$ in \eqref{HhA}. Given a finite projective left module $\cP$ over a
unital noncommutative ring $R$, there is a notion of trace of an endomorphism,
and of rank element, given by the trace of the identity, \cite{Hattori}. Consider 
the identification $$\vartheta: \cP^\vee \otimes_R \cP \stackrel{\simeq}{\to} \Hom_R(\cP,\cP),$$
with $\cP^\vee=\Hom_R(\cP,R)$ and $\vartheta(\eta\otimes x)(y)=\eta(y)x$. Also consider
the morphism $$\pi: \cP^\vee \otimes_R \cP \to HH_0(R)=R/[R,R],$$
with $[R,R]$ the additive commutators subgroup of $R$  
and $\pi(\eta\otimes x)=\eta(x)$ mod $[R,R]$.

\begin{defn}\label{tracef}{\rm
Given a finite projective left module $\cP$ over a unital noncommutative ring $R$,
the trace of an endomorphism $f\in \Hom_R(\cP,\cP)$ 
is the element  in $HH_0(R)$ obtained by setting $\Tr_R(f):=\pi(\vartheta^{-1}(f))$.
}\end{defn}

\smallskip

In the case where $R=M_n(D)$ is a matrix algebra over a division algebra
over a field $\K$, an element $X\in M_n(D)$ is in the commutator $[M_n(D),M_n(D)]$
iff its trace is in $[D,D]$.  
Moreover (Corollary~13.6 of \cite{Lam}) $D$ is generated as a division algebra 
by $Z(D)$ and all the commutators in $[D,D]$. If $D$ is central, $Z(D)=\K$, then
the trace elements can be seen as scalars in $\K$. More generally, the trace
elements determine scalars in the field extension $Z(D)$ of $\K$. 
In particular, for $f={\rm id}$ the
identity, $\Tr_R({\rm id})=:r_R(\cP)$ is the rank element of $\cP$. If $\cP = R e$ with $e$
an idempotent in $R$, then $r_R(\cP)$ is the class of $e$ in $HH_0(R)$ and it is independent
of the choice of idempotent: if $Ae\simeq A e' \simeq \cP$ then $e \sim e'$ in $HH_0(R)$,
see \cite{Hattori} for this and more general cases.  This rank element is known as
the Hattori--Stallings rank and the more general traces as in Definition~\ref{tracef} are
usually referred to as the Hattori--Stallings trace.

\smallskip

Thus, for our purposes, we summarize the above in the following statement.

\begin{lem}\label{rankel}
Let $A$ be a semisimple finite dimensional algebra over a number field $\K$
and $\cE$ a finite projective (left) $A$-module (see Remark~\ref{finprojcond}).
The rank element $r_A=r_A(\cE)$ is the class in $HH_0(A)=A/[A,A]$ given by
the trace of the identity, $\Tr_R({\rm id})=:r_A(\cE)$. A choice of an archimedean
embedding $\sigma: \K \hookrightarrow \C$ (or $\R$) and a compatible
embedding $\tilde\sigma$ of the extension $\bL$ of $\bK$ generated by the centers $Z(A_i)$
of the simple components $A_i$ of $A$ determine an associated rank 
$r_{A,\sigma}=\tilde\sigma(r_A)\in\C$ (or $\R$).
\end{lem}

\proof
The center $Z(A)$ of a semisimple algebra is a finite direct product of field extensions of $\K$, 
$Z(A)=\oplus_i \bL_i$, with $A=\oplus_i M_{n_i}(D_i)$ and $\bL_i=Z(D_i)=HH_0(M_{n_i}(D_i))$. 
The rank element $r_A \in HH_0(A)$ is then given by a collection of elements $r_{A,i}\in \bL_i$.  
Given an embedding $\sigma: \K \hookrightarrow \C$ (or into $\R$ for a real place), let 
$\tilde\sigma:\bL \hookrightarrow \C$ (or $\R$) be an extension of $\sigma$ to an embedding of the
smallest extension $\bL$ of $\bK$ that contains all the $\bL_i$. Then setting
$\sigma(r_A):= \sum_i \tilde\sigma(r_{A,i})$ assigns a complex (or real) number to 
the rank element $r_A\in HH_0(A)$.
\endproof

\smallskip

\smallskip
\subsection{Norms and volumes}\label{VolSec1}

There are different possible ways of associating to a finite dimensional normed
space of dimension $N$ a volume form (and $k$-volume forms for $k\leq N$):
some of these, such as Busemann volume, Holmes--Thompson volume,
Gromov mass, etc.~are summarized in \cite{AlPaiva}. We discuss
briefly the case of the Gromov mass, to illustrate what conditions need
to be satisfied for a good notion of volume. 

\smallskip

Let $(V,\|\cdot \|)$ be a normed space of dimension $N$. The  Gromov
$N$-mass is defined as a function $\mu_N: \Lambda^N V \to \R_+$
$$ 
\mu_N(a):=\inf \{ \prod_{i=1}^N \| v_i \| \,|\, v_1\wedge \cdots \wedge v_N =a \}.
$$

\smallskip

In order to define in a similar way $k$-volumes for $k< N=\dim V$, let
$\Lambda^k_s V \subset \Lambda^k V$ be the subset consisting of the
``simple" $k$-vectors, that is, those of the form $v_1\wedge \cdots \wedge v_k$,
with $v_i \in V$. In order to define a $k$-volume, it suffices to assign a
volume to elements of $\Lambda^k_s V$. Thus, one defines a $k$-volume
as a homogeneous continuous function $\mu_k: \Lambda^k_s V \to \R_+$ 
with $\mu_s(\lambda a)=|\lambda|\, \mu_s(a)$ for $\lambda\in \R^*$. The
analog of the Gromov mass $\mu_N$ is given by
$$
\mu_k(a) := \inf \{ \prod_{i=1}^k \| v_i \| \,|\, v_1\wedge \cdots \wedge v_k =a \}.
$$

\smallskip

In this notion of $k$-volumes, the constrained optimization of
the norms is designed to avoid redundancies that would make
the norm unnecessarily large, such as adding to a vector in a basis  
a large multiple of another basis vector, and also to avoid changes to
the norm due to an arbitrary scaling of a basis. The problem
simplifies if the normed vector space $V$ is endowed with a more rigid
structure, given by the choice of a lattice $V_\Z \subset V$. In this
case, if we want to assign $k$-volumes to elements of $\Lambda^k_s V_\Z$,
we can choose a different form of optimization of the norms by a priori
selecting a ``shortest basis" $\{ v_1, \ldots, v_N \}$ for the lattice $V_\Z$.
One can then define
\begin{equation}\label{mukprod}
 \mu_k (v_{i_1}\wedge\cdots \wedge v_{i_k}) := \prod_{\ell=1}^k \| v_{i_\ell} \|. 
\end{equation} 
for the basis elements $\{ v_{i_1}\wedge\cdots \wedge v_{i_k} \}$ of the $\Z$-module
$\Lambda^k V_\Z$. The shortest basis for a lattice is the basis with the smallest
possible orthogonality defect
\begin{equation}\label{defect}
 \delta(\{ v_i \}) =\frac{\prod_{i=1}^N \| v_i \|}{d(V_\Z)}, 
\end{equation} 
where $d(V_\Z)$ is the lattice constant, which is equal to $\det(B)$,
the determinant of the matrix $B=(v_1,\ldots,v_N)$ formed by the vectors $v_i$,
in the case where $V_\Z$ is of rank $N$. For rank $k\leq N$, one can still define 
the defect in the same way with $d(V_\Z)=\sqrt{\det(B^* B)}$, which measures the
Euclidean $k$-volume of the fundamental parallelepiped determined by the basis $\{ v_i \}$. 
The defect is therefore the ratio between the volume defined as in $\mu_k (v_{i_1}\wedge\cdots \wedge v_{i_k})$
and the Euclidean volume, and $\mu_k$ is computed on the choice of basis where this
discrepancy is as small as possible. A lattice reduction algorithm that approximates
the shortest basis for a lattice is developed in \cite{LLL}. 

\smallskip

Note that the quantity $d(V_\Z)=\sqrt{\det(B^* B)}$ has itself an interpretation in terms of
heights, since it is the archimedean part of the height of the matrix $B$ seen as a point in
a Grassmannian, see Remark~2.8.7 of \cite{BoGu}.

\smallskip

\begin{defn}\label{VolLattice}{\rm 
Let $\Lambda\subset V$ be a lattice of rank $k$ inside a $\K$-vector space of
dimension $N\geq k$, namely a finitely generated torsion free $\cO_\K$-module
such that $\K \Lambda$ spans a $k$-dimensional subspace of $V$. Given a norm
$\|\cdot \|$ on $V$, the $k$-volume of $\Lambda$ is 
\begin{equation}\label{kvolL}
{\rm vol}_k(\Lambda):= \mu_k (v_1 \wedge\cdots \wedge v_k)=\prod_{\ell=1}^k \| v_\ell \|,
\end{equation}
where $\{ v_1, \ldots, v_k \}$ is a shortest basis for $\Lambda$. }
\end{defn}

\smallskip

\begin{lem}\label{volOA}
Let $(A,\cO_A)$ be a pair of a semisimple finite
dimensional algebra $A$ over a number field $\K$ and an order $\cO_A$.
Let $\sigma$ be an archimedean place of $\K$ with $\K_\sigma$ equal
to $\R$ or $\C$ for real and complex places, respectively and let
$h_A: A_\sigma\times A_\sigma \to A_\sigma$ be a hermitian structure.
Let $\cE$ be a (left) $\cO_A$-module. The notion of
volume of lattices of Definition~\ref{VolLattice} and the rank element $r_A(\cE)$
of Lemma~\ref{rankel} determine the quantity ${\rm vol}_{h_A,\sigma}(\cO_A)^{r_{A,\sigma}}$ of
\eqref{HhA}.
\end{lem}

\proof As shown in \cite{Borek1} all hermitian structures on $A_\sigma=A\otimes_\K \K_\sigma$
are standard in the sense of Definition~\ref{standardh}, hence there is an element $\beta\in A^*$
such that $h_A(x,y)=x^* \beta^* \beta y$, and $h_A(x,x)$ is a positive element for all $x\in A_\sigma$.
Thus, we have $\| x \|^2 :=\Tr(h_A(x,x))  \geq 0$ for all $x\in A_\sigma$. By Definition~\ref{orderOA},
an order $\cO_A$ in $A$ is a $\cO_\K$-lattice with $\K \cO_A =A$ as a $\K$-vector space, hence
the norm determined by $h_A$ also determines a volume, as in \eqref{kvolL}, of the lattice $\cO_A$
with $k=\dim_\K A$, the dimension as $\K$-vector space. With $r_{A,\sigma}=\tilde\sigma(r_A(\cE))$
as in Lemma~\ref{rankel}, we then obtain the expression in \eqref{HhA} that generalizes to our
setting the height function \eqref{BorekH} used in \cite{Borek1}.
\endproof

\smallskip

We will see in the following how to obtain also 
a good definition of the volumes ${\rm vol}_{h_A,\sigma}(\cE\cap \cO_A^n)$, 
suitable for extending the definition of the height function as in \eqref{HhA}
to our notion of arithmetic vector bundles.

\smallskip
\subsection{Volumes, traces and heights}\label{VolSec}

We look here more explicitly at the height function defined in \eqref{HEhAB} by
writing more explicitly the traces and volume forms on the bimodules. We will assume
here that hermitian structures as standard as in Definition~\ref{standardh}.

\smallskip

\begin{prop}
Consider algebras of the form $$ A=\oplus_i M_{n_i}(D_i) \ \ \text{ and } \ \ 
B=\oplus_j M_{m_j}(D'_j)$$ where $D_i,D'_j$ are division algebras over the
number field $\K$. We also assume that an order $\cO_A$ (respectively, $\cO_B$)
has been chosen. Let $\cE$ be an $\cO_A$-$\cO_B$ bimodule. 
For every choice of a (standard) hermitian structure $(h_A, h_B)$ and archimedean
place $\sigma$ of $\K$, there are associated volumes 
$vol_{h_A,\sigma}(j_A(\cE)\cap \cO_A^{n_A})$ and
$vol_{h_B,\sigma}(j_B(\cE)\cap \cO_B^{n_B})$, where $j_A: \cE \hookrightarrow A^{n_A}$
and $j_B: \cE \hookrightarrow B^{n_B}$ realize the projective $A$-module (respectively, $B$-module)
as a summand of a free module.
\end{prop}

\proof
For semisimple algebras all left (respectively, right) modules are
semisimple (\cite{Lam}, Theorem~2.5), hence we can consider
$A$-$B$ bimodules that are of the form
\begin{equation}\label{bimodDDprime}
 \cE = \oplus_{i,j} D_i^{n_i N_i} \otimes {D'_j}^{m_j M_j}, 
\end{equation} 
with the obvious left and right action of the respective matrix algebras. 

\smallskip

Standard hermitian structures $h_A: \cE \times \cE \to A$ and $h_B: \cE\times \cE \to B$
are obtained as follows.

\smallskip

A choice of elements $\beta_{A,i}\in M_{n_i}(D_i)^*$ and $\beta_{B,j}\in M_{m_j}(D_j')^*$.
determines hermitian forms $h_{A,i}: D_i^{n_i} \times D_i^{n_i} \to D_i$
and $h_{B,j}: {D'_j}^{m_j} \times {D'_j}^{m_j} \to D'_j$ by
$$ h_{A,i}(x,y):= x \beta_{A,i} \beta_{A,i}^* y^* \ \ \ \text{ and } \ \ \ h_{B,j}(u,v):= u^* \beta_{B,j}^* \beta_{B,j} v $$
where $x=(x_1,\ldots, x_{n_i})$ in $D_i^{n_i}$ (as a column vector) with $x^*$ the transpose (row vector) $(x_1^*,\ldots, x_{n_i}^*)$, and similarly for $h_{B,j}$.
These extend similarly to hermitian forms $h_{A,i}: D_i^{n_i N_i} \times D_i^{n_i N_i} \to D_i$
and $h_{B,j}: {D'_j}^{m_j M_j} \times {D'_j}^{m_j M_j} \to D'_j$. 

\smallskip

Here we are using the notation $*$ both for an involution on $M_n(D)$ and for an involution on $D$,
assuming compatible involutions on both. For example, if 
$M_n(D)$ is a central simple algebra, then it admits an involution that is the identity on the field $\K$
(involution of the first kind) iff the division algebra $D$ admits such an involution, with the
involution on $M_n(D)$ given by the involution on $D$ applied to all entries, combined with matrix
transposition. 

\smallskip

These hermitian structures on the $D_i^{n_i N_i}$ and ${D'_j}^{m_j M_j}$ determine
hermitian structures $h_A: \cE \times \cE \to A$ and $h_B: \cE\times \cE \to B$ as 
shown in Lemma~\ref{Hbilem}. In turn, for any archimedean place 
$\sigma$ of $\K$, these hermitian forms determine hermitian structures 
$h_A: \cE_\sigma\times \cE_\sigma \to A_\sigma$, and similarly for $h_B$.

\smallskip

Let $\{ e_{i,k} \}$ denote a basis of $D_i^{n_i N_i}$ as a left
$M_{n_i}(D_i)$-module and similarly let $\{ e_{j,k'}' \}$ denote a 
basis of ${D'_j}^{m_j M_j}$ as a right $M_{m_j}(D'_j)$-module. Let $\{ v'_{j,\ell'} \}$
be a basis of ${D'_j}^{m_j M_j}$ as a $\K$-vector space and $\{ v_{i,\ell} \}$
be a basis of $D_i^{n_i N_i}$ as a $\K$-vector space, so that, as in Lemma~\ref{Hbilem},
we can consider $u_{i,j;k,\ell'}=e_{i,k}\otimes v'_{j,\ell'}$ and $u'_{i,j; \ell, k'}=v_{i,\ell}\otimes e_{j,k'}'$
as bases, respectively, of $\cE$ as a left $A$-module and as a right $B$-module, so that
we can write elements $\xi\in \cE$ in the form
$\xi=\sum_{i,j,k,\ell'} a_{i,j;k,\ell'} u_{i,j;k,\ell'}$ with $a_{i,j;k,\ell'}\in M_{n_i}(D_i)$ or
in the form $\xi=\sum_{i,j,\ell,k'} u'_{i,j; \ell, k'} b_{i,j; \ell, k'}$ with $b_{i,j; \ell, k'}\in M_{m_j}(D'_j)$.

\smallskip

Since $\cE$ is a finite projective module, both as a left $A$-module and as a right $B$-module
(see Remark~\ref{finprojcond}), we can use a choice of bases to determine embeddings
$j_A: \cE \hookrightarrow A^{n_A}$ and $j_B: \cE \hookrightarrow B^{n_B}$, which realizes
$\cE$ as $\cE\simeq \cO_A^{n_A} e_A$ with an idemponent $e_A\in M_{n_A}(A)$ with
components $(e_A)_{ij}$ determined by the basis elements, and similarly for the $B$-module
structure. 

\smallskip

If the bimodule $\cE$ is arithmetic, namely it is obtained as an $A$-$B$ bimodule by change
of coefficients $\cE=A\otimes_{\cO_A} \cE_\cO \otimes_{\cO_B} B$ from an 
$\cO_A$-$\cO_B$ bimodule $\cE_\cO$, then it is possible to choose the bases 
in such a way that they are in $\cE_\cO$, with the $\{ e_{i,k} \}$ and $\{ e_{j,k'}' \}$ giving
$\cO_A$ and $\cO_B$ bases, respectively, and the $\{ v_{i,\ell} \}$ and $\{ v'_{j,\ell'} \}$ 
bases of $\cO_\K$-lattices, so that the resulting $\{ u_{i,j;k,\ell'} \}$ and 
$\{ u_{i,j;k,\ell'} \}$ give bases of $\cE_\cO$ as a left and right
$\cO_A$ and $\cO_B$ module, respectively. Let $\cB, \cB'$ denote the set of all such bases
and let $u_{\min}, u_{\min}'$ be shortest basis
$$ \delta(u_{\min}) =\inf_{u\in \cB} \delta(u)\ \ \  \text{ and } \ \ \ \delta(u'_{\min}) =\inf_{u'\in \cB'} \delta(u'), $$
with $\delta$ as in \eqref{defect}, where the norm $\|\cdot\|=\Tr(h_A(\cdot,\cdot))^{1/2}$ (or $h_B$, respectively)
and the lattice constant $d$ is computed as the Euclidean volume as $\Z$-lattices. 

\smallskip

We can then consider $j_A(\cE_\cO)\cap \cO_A^{n_A}$
and $j_B(\cE_\cO) \cap \cO_B^{n_B}$. These can be viewed as $\cO_A$-module (respectively,
$\cO_B$-module) and as $\cO_\K$-modules. Let $k_A$, $k_B$ denote, respectively, their
ranks as $\cO_\K$-modules. 

\smallskip

We can then assign a volume to $j_A(\cE_\cO)\cap \cO_A^{n_A}$
and $j_B(\cE_\cO) \cap \cO_B^{n_B}$ as in \eqref{kvolL}, computed with respect 
to a shortest basis, in the sense discussed above.
\endproof

Thus, we obtain a well defined height functions
\begin{equation}\label{relHE}
 \bar H_{\cO_A,h_A,\sigma}(\cE)=\frac{{\rm vol}_{h_A,\sigma}(j_A(\cE)\cap \cO_A^{n_A})}{{\rm vol}_{h_A,\sigma}(\cO_A)^{\sigma(r_A)}} \ \ \  \text{ and } \ \ \  \bar H_{\cO_B,h_B,\sigma}(\cE)=\frac{{\rm vol}_{h_B,\sigma}(j_B(\cE)\cap \cO_B^{n_B})}{{\rm vol}_{h_B,\sigma}(\cO_B)^{\sigma(r_B)}}
 \end{equation}
 that can be used in the height function of Definition~\ref{HEhABdef}.

\smallskip
 \subsection{Multiplicative property}
 
 The issue here is that, while the height function \eqref{relHE}
 assigns to bimodules an ``energy" as we want, it only sees bimodules as forming a {\em set},
 while we also want to keep track of the structure of our functor of points as a {\em category}.
 This means that we want a height function that is in a natural way compatible
 with the composition operation given by the tensor product of bimodules, while
 still being based on an appropriate notion of volumes. To obtain a simple
 setting that satisfies this property, we will restrict the class of bimodules and
 the choice of basis used to compute the height.
 
 \smallskip
 
 Arithmetic hermitian bimodules behave in the following way under
 tensor product.
 
 \begin{lem}\label{tensprodE}
The tensor product $(\cE,h_A,h_C)$ in ${}_{(A,\cO_A)}\cH_{(C,\cO_C)}$
of arithmetic hermitian bimodules $(\cE',h'_A,h'_B)$ in
${}_{(A,\cO_A)}\cH_{(B,\cO_B)}$ and $(\cE'',h''_B,h''_C)$ in
${}_{(B,\cO_B)}\cH_{(C,\cO_C)}$ is given by
$\cE=\cE'\otimes_{\cO_B} \cE''$ with the hermitian structures
$$ h_A(x'\otimes x'',y'\otimes y''):=h'_A(x' \, h''_B(x'',y''), y') $$
$$ h_C(x'\otimes x'',y'\otimes y''):=h''_C(x'', h'_B(x',y')\, y''). $$
\end{lem}

\proof To prove this Lemma, one needs to show that $(\cE,h_A,h_C)$ satisfies all the properties of Definition~\ref{Hbimods}. By construction, $\cE=\cE'\otimes_{\cO_B} \cE''$ is an  $\cO_A-\cO_C$ bimodule since $\cE'$ is an $\cO_A$-left module and $\cE''$ is an $\cO_C$-right module, and those two actions are compatible. In addition, Remark~\ref{finprojcond} implies that $A\otimes_{\cO_A} \cE$ and $\cE\otimes_{\cO_C} C$  are finite projective
as $A$-module and $C$-module, respectively.

For any archimedean place $\sigma$ of $\K$, one needs to show that the
bimodule $\cE_\sigma:=A_\sigma \otimes_{\cO_A} 
\cE \otimes_{\cO_C} C_\sigma$ endowed with the pair of hermitian 
structures $h=(h_A,h_B)$, as defined above, with $h_A: \cE_\sigma \times \cE_\sigma \to A_\sigma$
and $h_C: \cE_\sigma \times \cE_\sigma \to C_\sigma$ satisfying all the conditions in Definition \ref{Hbimods}:   
\begin{itemize}
\item For all $a\in A_\sigma$ and $(x'\otimes x'',y'\otimes y'') \in \cE_\sigma$, the identities 
\begin{align*}
h_A(a (x'\otimes x''), y'\otimes y'')&=h'_A(ax' \, h''_B(x'',y''), y')\\
&=a \, h'_A(x' \, h''_B(x'',y''), y') \\
&= a \,h_A( x'\otimes x'', y'\otimes y'')
\end{align*}
and 
\begin{align*}
h_A(x'\otimes x'', a(y'\otimes y''))&=h'_A(x' \, h''_B(x'',y''), ay')\\
&= \, h'_A(x' \, h''_B(x'',y''), y')\, a^* \\
&=  \,h_A( x'\otimes x'', y'\otimes y'')\, a^*
\end{align*}
where the second last equalities hold since $h'_A$ satisfies conditions in Definition \ref{Hbimods}.
\item For all $x'\otimes x'' \in \cE_\sigma$, the element $h_A(x'\otimes x'',x'\otimes x'')$ is a positive element
in $A_\sigma$ since there is a $\beta_{B}\in B_{\sigma}^*$ such that
\begin{align*}
h_A(x'\otimes x'', x'\otimes x'')&=h'_A(x' \, h''_B(x'',x''), x')\\
&=  h'_A(x' \, x'' \beta_{B} \beta_{B}^* x''^*, x')\\
&=  h'_A(x' \, x'' \beta_{B} , x'x'' \beta_{B})
\end{align*}
and $h'_A(x' \, x'' \beta_{B} , x'x'' \beta_{B})$ is positive element in $A_\sigma$ given that $ h'_A$ satisfies condition in Definition \ref{Hbimods}.
\end{itemize}
And similarly reasoning holds for $h_C$, finishing the proof. 
\endproof

\smallskip

Let $A$ be a semisimple finite dimensional algebra over a number field $\K$. Then
by Wedderburn theorem $A=\oplus_i M_{n_i}(D_i)$ with division algebras $D_i$ over $\K$.
The center $Z(A)$ is then a direct sum of number fields $\bL_i$ that contain $\K$.

\smallskip

\begin{defn}\label{bimodZ}
Let $\cA_\K(Z)$ denote the set of all semisimple finite dimensional algebras $A$ over $\K$
such that $Z(A)\simeq Z$ with $Z$ a fixed $Z=\oplus_i \bL_i$.
Let $\cH(\K,Z)$ denote the set of $A$-$B$ bimodules $\cE$ satisfying the following properties
\begin{itemize}
\item The algebras $A,B$ are in $\cA_\K(Z)$;
\item $\cE$ is an arithmetic $(A,\cO_A)$-$(B,\cO_B)$ hermitian bimodule, for some order $\cO_A$ in $A$
and $\cO_B$ in $B$, with hermitian structures $(h_A,h_B)$;
\item There are bases $\{ x_i \}$ and $\{ y_j \}$ of $\cE$ as a left $A$-module and a right $B$-module
respectively, with the property that $h_A(x_i,x_i) \in Z$ and $h_B(y_j,y_j)\in Z$;
\item As a $Z$-$Z$ bimodule $\cE$ is a symmetric bimodule, namely $\lambda \xi = \xi \lambda$ for
all $\xi\in \cE$ and all $\lambda\in Z$.
\end{itemize}
Let $\cB_A(\cE)$ and $\cB_B(\cE)$ be, respectively, the set of 
all bases $\{ x_i \}$ and bases $\{ y_j \}$ that satisfy the condition above
that $h_A(x_i,x_i) \in Z$ and $h_B(y_j,y_j)\in Z$ and such that $x_i, y_j \in \cE_\cO$, for all $i,j$.
\end{defn}

Note that the sum $\sum_i h_A(x_i,x_i)$ is the trace of the identity endomorphism
and is in $Z$ (identified with $HH_0(A)$) for an arbitrary choice of basis, and 
similarly for $\sum_j h_B(y_j,y_j)$.
The condition that $h_A(x_i,x_i) \in Z$ and $h_B(y_j,y_j)\in Z$ is verified, for
instance when $\cE= A^{n_A} p$, for a projector $p\in M_{n_A}(Z(A))$ with diagonal
and similarly $\cE =p' B^{n_B}$ with $p'\in M_{n_B}(Z(B))$.

\smallskip
 
 We can then considered a $2$-category $\cH(\K,Z)$ as follows.

\begin{lem}\label{2catHmods}
Let $\K$ be a number field and let $Z=\oplus_i \bL_i$ be a given sum of number field extensions $\bL_i$ of $\K$.
The following data form a $2$-category $\cH(\K,Z)$: 
\begin{itemize}
\item Objects:  pairs $(A,\cO_A)$ of a semisimple finite dimensional algebra $A$ in $\cA_\K(Z)$
with a $\cO_\K$-order $\cO_A$ in $A$;
\item $1$-morphisms ${\rm Mor}((A,\cO_A),(B,\cO_B))$: arithmetic $(A,\cO_A)$-$(B,\cO_B)$
hermitian bimodules $(\cE,h_A,h_B)$ with $\cE$ in $\cH(\K,Z)$;
\item compositions of $1$-morphisms $(\cE,h'_A,h'_B)$ and $(\cF, h''_B, h''_C)$ is
$$(\cE'\otimes_B \cE'', h_A, h_C),$$ with $h_A$, $h_C$ as in Lemma~\ref{tensprodE};
\item $2$-morphisms: morphisms $\phi: (\cE,h_A,h_B)\to (\tilde\cE,\tilde h_A,\tilde h_B)$ of hermitian bimodules, namely
morphisms of bimodules that preserve the hermitian forms, $\tilde h_A(\phi(x),\phi(y))=h_A(x,y)$ and
$\tilde h_B(\phi(x), \phi(y))=h_B(x,y)$, with the property that of $\{ x_i \}$ and $\{ y_j \}$ are bases of $\cE$
$h_A(x_i,x_i) \in Z$ and $h_B(y_j,y_j)\in Z$, then $\{ \phi(x_i) \}$ and $\{ \phi (y_j) \}$ can be completed
to bases of $\cF$ with the same property;
\item horizontal composition: tensor product of morphisms of hermitian bimodules 
\item vertical composition: composition of morphisms of bimodules. 
\end{itemize}
\end{lem}

\smallskip

\proof If $\{ x_i \}$ and $\{ y_j \}$ are bases of $\cE$ as left $A$-module and a right $B$-module,
satisfying $h'_A(x_i,x_i) \in Z$ and $h'_B(y_j,y_j)\in Z$, and $\{ u_\ell \}$ and $\{ v_k \}$ are bases of
$\cF$ as a left $B$-module and a right $C$-module, also satisfying $h''_B(u_\ell,u_\ell)\in Z$ and
$h''_C(v_k,v_k)\in Z$, then $\{ x_i \otimes u_\ell \}$ and $\{ y_j \otimes v_k \}$ are bases for
$\cE\otimes_B \cF$ as a left $A$-module and a right $C$-module satisfying 
\begin{equation}\label{hAhBprod}
 h_A( x_i \otimes u_\ell, x_i \otimes u_\ell)= h_A'(x_i h''_B(u_i,u_i), x_i) =h''_B(u_i,u_i) h_A'(x_i,x_i) \in Z, 
\end{equation} 
where the second equality uses the symmetric property of $\cE$ as a $Z$-$Z$ bimodule, and similarly
\begin{equation}\label{hBhCprod}
 h_C( y_j \otimes v_k , y_j \otimes v_k )=h''_C(v_k, h'_B(y_j,y_j) v_k)=h''_C(v_k,v_k) h'_B(y_j,y_j)  \in Z,
\end{equation} 
similarly using the symmetry of $\cF$ as $Z$-$Z$ bimodule. The product $\cE\otimes_B \cF$ is also
symmetric as $Z$-$Z$ bimodule. Thus, composition of $1$-morphisms is well defined in $\cH(\K,Z)$. 
The properties of the vertical and horizontal compositions of $2$-morphisms are similarly verified.
\endproof

\smallskip

\begin{rem}\label{pointedFunctPts}{\rm
Our (pointed) noncommutative functor of points $X^{nc}_{(A,\cO_A),\star}$ can then
be thought of as assigning to $(A,\cO_A)$ with $A\in \cA_\K(Z)$ the collection of 
all the arrows ($1$-morphisms) in $\cH(\K,Z)$ with source $(A,\cO_A)$.}
\end{rem}

\smallskip

We now define a height function as follows.

\begin{defn}\label{ptdH}
Suppose given an $(A,\cO_A)$-$(B,\cO_B)$ arithmetic hermitian bimodule $(\cE,h_A,h_B)$ in $\cH(\K,Z)$,
with $A,B \in \cA_\K(Z)$, and an archimedean place $\sigma: \K \to \K_\sigma$.
The heights of $(\cE,h_A,h_B)$ are given by 
\begin{equation}\label{HptdAB}
\begin{array}{rl}
H_{h_A,\sigma}(\cE):= & \inf_{\{ x_i \} \in \cB_A(\cE)} \prod_i \tilde\sigma(h_A(x_i,x_i)) \\[2mm]
H_{h_B,\sigma}(\cE):= & \inf_{\{ y_j \} \in \cB_B(\cE)} \prod_j \tilde\sigma(h_B(y_j,y_j)) ,
\end{array}
\end{equation}
with $\tilde\sigma$ an embedding of $Z$ in $\C$ that restricts to the embedding $\sigma: \K \hookrightarrow \C$.
Correspondingly, we define a height function as in Definition~\ref{HEhABdef} with
 \begin{equation}\label{HEZ}
  H(\cE,h_A,h_B) := \prod_\sigma \frac{H_{h_B,\sigma}(\cE)}{H_{h_A,\sigma}(\cE)} .
 \end{equation}
\end{defn}

\smallskip

\begin{lem}\label{multiplHAB}
The height function defined as in \eqref{HEZ} satisfies
\begin{equation}\label{heighttensor}
H(\cE\otimes_B \cF,h_A,h_B) = H(\cE, h'_A, h'_B) \cdot H(\cF, h''_B, h''_C),
\end{equation}
for $(\cE,h'_A, h'_B)$ an $(A,\cO_A)$-$(B,\cO_B)$ hermitian bimodule and $(\cF, h''_B, h''_C)$ an
$(B,\cO_B)$-$(C,\cO_C)$ hermitian bimodule, both in $\cH(\K,Z)$, and with $h_A,h_B$ as in 
Lemma~\ref{tensprodE}.
\end{lem}

\proof For any bases $\{ x_i \}$ and $\{ y_j \}$ of $\cE$ in $\cB_A(\cE)$ and $\cB_B(\cE)$,
respectively, and bases $\{ u_\ell \}$ and $\{ v_k \}$ of $\cF$ in $\cB_B(\cF)$ and $\cB_C(\cF)$,
by \eqref{hAhBprod} and \eqref{hBhCprod}, we obtain a basis $\{ x_i \otimes u_\ell \}$
in $\cB_A(\cE \otimes_B \cF)$ and a basis $\{ y_j \otimes v_k \}$ in $\cB_C(\cE\otimes_B \cF)$. 
Moreover, also by \eqref{hAhBprod} and \eqref{hBhCprod} 
$$ \prod_{i,\ell} \tilde\sigma( h_A(x_i \otimes u_\ell, x_i \otimes u_\ell))^{1/2} 
=\prod_i \tilde\sigma(h'_A(x_i,x_i))  \cdot \prod_\ell \tilde\sigma(h''_B(u_\ell,u_\ell)) $$
$$ \prod_{j,k} \tilde\sigma( h_C(y_j \otimes v_k, y_j \otimes v_k)) =
\prod_j \tilde\sigma( h'_B(y_j , y_j)) \prod_k \tilde\sigma( h''_C(v_k, v_k)), $$
hence minimizing over the choice of basis, 
$$ H_{h_A,\sigma}(\cE\otimes_B \cF) = H_{h'_A,\sigma}(\cE) H_{h''_B,\sigma}(\cF) $$
$$ H_{h_C,\sigma}(\cE\otimes_B \cF) = H_{h'_B,\sigma}(\cE) H_{h''_C,\sigma}(\cF), $$
so that
$$ H(\cE\otimes_B \cF,h_A,h_C) = \frac{H_{h_C,\sigma}(\cE\otimes_B \cF)}{H_{h_A,\sigma}(\cE\otimes_B \cF)}=
\frac{H_{h'_B,\sigma}(\cE)}{ H_{h'_A,\sigma}(\cE) }  \cdot \frac{H_{h''_C,\sigma}(\cF)}{H_{h''_B,\sigma}(\cF)}. $$
\endproof

\smallskip
\subsection{Dynamics generated by the height function}

We explain more clearly in this subsection in what sense the height function
introduced in the previous subsection makes our functor of points 
$(B,\cO_B) \mapsto X^{nc}_{(A,\cO_A)}(B,\cO_B)$
dynamical. To this purpose we take the point of view of quantum statistical mechanics
and we define an algebra of observables over $X^{nc}_{(A,\cO_A)}(B,\cO_B)$
with a time evolution generated by the height function. The main idea is very similar
to the algebra and dynamics constructed in \cite{CoMa2}, associated to the space
of $\Q$-lattices with the commensurability relation, but the construction we present here 
is based on the convolution algebras of semigroupoids (small categories) and of
$2$-categories introduced in \cite{MaZa}.

\smallskip

As we discussed in the introduction to this paper, we consider the
height as a kind of ``energy functional" on the ``points" (here represented by
bimodules) of a noncommutative space. We mentioned that this action functional 
makes our functor of points ``dynamical". To make this idea 
more precise, we associate to the $2$-category $\cH(\K,Z)$ of bimodules whose
$1$-morphisms describe our functor of points, an algebra of observables and a time 
evolution determined by the action functional given by the height function \eqref{HEZ}.

\smallskip

As shown in \cite{MaZa} there are two convolution algebras $\cA^{(1)}_{\cH(\K)}$ and 
$\cA^{(2)}_{\cH(\K)}$ associated to a $2$-category $\cH(\K)$. The first is based on $1$-morphisms
and their composition, and the second is based on $2$-morphisms and their horizontal and
vertical compositions. 

\smallskip

The algebra $\cA^{(1)}_{\cH(\K,Z)}$ is given by finitely supported
functions on the set of $1$-morphisms of $\cH(\K,Z)$, with the convolution product
\begin{equation}\label{convol1}
 (f_1\star f_2) (\Phi)=\sum_{\Phi=\Phi''\star \Phi'} f_1(\Phi') f_2(\Phi''), 
\end{equation} 
for $\Phi=(\cE,h_A,h_B)$ in ${}_{(A,\cO_A)}\cH_{(B,\cO_B)}$ equal to the tensor product of
$\Phi'=(\cE',h_A,h_B)$ in ${}_{(A,\cO_A)}\cH_{(B,\cO_B)}$ 
and $\Phi''=(\cE'',h_b,h_C)$ in ${}_{(B,\cO_B)}\cH_{(C,\cO_C)}$, as in Lemma~\ref{tensprodE}, 
with the composition $\Phi''\star \Phi'$ of $1$-morphisms given by the tensor product of
hermitian bimodules. This algebra is associative but noncommutative. 

\smallskip

The algebra $\cA^{(2)}_{\cH(\K,Z)}$ is similarly defined but based on the two composition
laws for $2$-morphisms. It is given by finitely supported functions on the set of
$2$-morphisms (morphisms of hermitian bimodules) with two product operations
$$ (f_1\circ f_2)(\phi)=\sum_{\phi=\phi''\circ \phi'} f_1(\phi') f_2(\phi'') $$
$$ (f_1\bullet f_2)(\phi)=\sum_{\phi=\phi''\bullet \phi'} f_1(\phi') f_2(\phi'') $$
where $\circ$ is the vertical composition of $2$-morphisms 
$$ \phi''\circ \phi' : (\cE,h_A,h_B)\stackrel{\phi'}{\rightarrow} (\cE', h_A', h_B') 
\stackrel{\phi''}{\rightarrow} (\cE'', h_A'', h_B'') $$
while $\phi''\bullet \phi'$ is the horizontal composition deternined by the
tensor product of hermitian bimodules. 

\smallskip

\begin{lem}\label{timeovA1}
The height function \eqref{HEZ} for hermitian bimodules in $\cH(\K,Z)$
determines a time evolution on the convolution algebra $\cA^{(1)}_{\cH(\K,Z)}$
of $1$-morphisms of $\cH(\K,Z)$.
\end{lem}

\proof We construct a $1$-parameter family of automorphisms of $\cA^{(1)}_{\cH(\K,Z)}$,
that is, a group homomorphism $\alpha: \R \to {\rm Aut}(\cA^{(1)}_{\cH(\K,Z)})$ by
taking
$$ \alpha_t(f)(\cE,h_A,h_B) = H(\cE,h_A,h_B)^{it} \, f(\cE,h_A,h_B) $$
Clearly, $\alpha_{t+s}=\alpha_t\circ \alpha_s$. We need to check that
the $\alpha_t$ are algebra homomorphisms
$$ \alpha_t(f_1\star f_2)(\cE,h_A,h_B)=(\alpha_t(f_1)\star \alpha_t(f_2))(\cE,h_A,h_B). $$
The left-hand-side is given by $H(\cE,h_A,h_B)^{it} \,\sum f_1(\cE',h_A',h_C') f_2(\cE'',h_C'',h_B'')$,
with the sum over all the decompositions of $(\cE,h_A,h_B)$ as tensor product of
$(\cE',h_A',h_C')$ and $(\cE'',h_C'',h_B'')$. The right hand side is given by
$$\sum H(\cE',h'_A,h'_B)^{it} f_1(\cE',h_A',h_C')\, H(\cE'',h''_A,h''_B)^{it} f_2(\cE'',h_C'',h_B'').$$
Lemma~\ref{multiplHAB} shows that the height function behaves multiplicatively
on a tensor product of bimodules, $H(\cE,h_A,h_C)=H(\cE',h'_A,h'_B) H(\cE'',h''_B,h''_C)$,
so that these two expressions agree. 
\endproof

\smallskip

Thus, the (non-normalized) height \eqref{HEZ} of hermitian bimodules in $\cH(\K,Z)$ 
makes the spaces $X^{nc}_{(A,\cO_A)}$
(of which $\cA^{(1)}_{\cH(\K)}$ represents the algebra of functions) dynamical
through the time evolution $\alpha_t$. Notice that this dynamics, generated by
a ratio of volumes on the two sides of a hermitian bimodule, is very similar to
the time evolution generated by the ratio of volumes of two commensurable
lattices in the case of the quantum statistical mechanical system of \cite{CoMa2}.
We can similarly define a time evolution on the algebra of $2$-morphisms $\cA^{(2)}_{\cH(\K,Z)}$
of $\cH(\K,Z)$, using a notion of relative height.

\smallskip

\begin{defn}\label{relheight}{\rm
The relative height $H(\phi)$ of a $2$-morphism $\phi: (\cE,h_A,h_B)\to (\cE', h_A', h_B')$
in $\cH(\K,Z)$ is defined as 
\begin{equation}\label{relH}
H(\phi) : = \frac{H(\tilde \cE, \tilde h_A, \tilde h_B)}{H(\cE,h_A,h_B)}. 
\end{equation}
}\end{defn}

\smallskip

\begin{lem}\label{Hphirel}
We can write \eqref{relH} equivalently in the form
$$ H(\phi)=\frac{\inf_{\{ y_r\}\in \cB_B(\tilde\cE)\smallsetminus \cB_B(\cE)} \prod_\sigma \prod_r \tilde\sigma(\tilde h_B(y_r, y_r))}{\inf_{\{ x_k\}\in \cB_A(\tilde\cE)\smallsetminus \cB_A(\cE)} \prod_\sigma \prod_k \tilde\sigma(\tilde h_A(x_k, x_k))}. $$
\end{lem}

\proof
Let us recall Definition \ref{ptdH}, which tells us that, 
$$
 H(\cE,h_A,h_B) := \prod_\sigma \frac{H_{h_B,\sigma}(\cE)}{H_{h_A,\sigma}(\cE)} $$
where 
\begin{equation*}
\begin{array}{rl}
H_{h_A,\sigma}(\cE):= & \inf_{\{ x_i \} \in \cB_A(\cE)} \prod_i \tilde\sigma(h_A(x_i,x_i)) \\[2mm]
H_{h_B,\sigma}(\cE):= & \inf_{\{ y_j \} \in \cB_B(\cE)} \prod_j \tilde\sigma(h_B(y_j,y_j)).
\end{array}
\end{equation*}
Then, by definition, we get 
\begin{align*}
H(\phi) :& = \frac{H(\tilde \cE, \tilde h_A, \tilde h_B)}{H(\cE,h_A,h_B)}\\
&= \frac{\prod_\sigma \frac{H_{\tilde h_B,\sigma}(\cE)}{H_{\tilde h_A,\sigma}(\cE)}}{\prod_\sigma \frac{H_{h_B,\sigma}(\cE)}{H_{h_A,\sigma}(\cE)}}\\
&= \frac{\big(\inf_{\{ \tilde y_r \} \in \cB_B( \tilde \cE)} \prod_\sigma \prod_r \tilde\sigma(\tilde h_B(\tilde y_r,\tilde y_r))\big) \big(\inf_{\{ x_i \} \in \cB_A(\cE)} \prod_\sigma \prod_i \tilde\sigma( h_A(x_i,x_i)))\big)}{\big(\inf_{\{ \tilde x_k \} \in \cB_A(\tilde \cE)} \prod_\sigma \prod_k \tilde\sigma(\tilde h_A(\tilde x_k,\tilde x_k))\big)\big(\inf_{\{ y_j \} \in \cB_B(\cE)} \prod_\sigma \prod_j \tilde\sigma(h_B(y_j,y_j))\big)}\\
&= \frac{\big(\inf_{\{ \tilde y_r \} \in \cB_B( \tilde \cE)} \prod_\sigma \prod_r \tilde\sigma(\tilde h_B(\tilde y_r,\tilde y_r))\big) \big(\inf_{\{ x_i \} \in \cB_A(\cE)}\prod_\sigma  \prod_i \tilde\sigma(\tilde h_A(\phi(x_i),\phi(x_i)))\big)}{\big(\inf_{\{ \tilde x_k \} \in \cB_A(\tilde \cE)} \prod_\sigma \prod_k \tilde\sigma(\tilde h_A(\tilde x_k,\tilde x_k))\big)\big(\inf_{\{ y_j \} \in \cB_B(\cE)} \prod_\sigma \prod_j \tilde\sigma(\tilde h_B(\phi(y_j), \phi(y_j)))\big)}\\
\end{align*}

 \endproof

\smallskip

\begin{lem}\label{A2timeev}
The relative non-normalized height $H(\phi)$ of \eqref{relH} determines a time evolution on the
algebra $\cA^{(2)}_{\cH(\K,Z)}$ of $2$-morphisms that is compatible with both
vertical and horizontal composition.
\end{lem}

\proof We define the time evolution as $\alpha_t(f)(\phi)=H(\phi)^{it}\, f(\phi)$.
To verify compatibility with the two product structures of $\cA^{(2)}_{\cH(\K,Z)}$,
consider a vertical composition $\phi''\circ \phi'$ with 
$\phi' : (\cE,h_A,h_B)\to (\cE', h_A', h_B')$ and $\phi'': (\cE', h_A', h_B')\to (\cE'', h_A'', h_B'')$. 
By \eqref{relH} we have $H(\phi''\circ \phi')=H(\phi') H(\phi'')$, hence we have
$$ \alpha_t(f_1\circ f_2)(\phi)=H(\phi)^{it} \sum_{\phi=\phi''\circ \phi'} f_1(\phi') f_2(\phi'') =
\sum_{\phi=\phi''\circ \phi'} \alpha_t(f_1)(\phi')\, \alpha_t(f_2)(\phi''). $$
In the case of a horizontal composition of $\phi': (\cE,h_A,h_B)\to (\tilde\cE,\tilde h_A,\tilde h_B)$
with $(\cE,h_A,h_B)$ the tensor product of $(\cE',h'_A,h'_B)$ and $(\cE'',h''_A,h''_B)$ and
$(\tilde\cE,\tilde h_A,\tilde h_B)$ the tensor product of $(\tilde \cE',\tilde h'_A,\tilde h'_B)$ and $(\tilde \cE'',\tilde h''_A,\tilde h''_B)$, as in Lemma~\ref{timeovA1}, with $\phi=\phi''\bullet \phi'$,  by Lemma~\ref{Hphirel} we have
$$ \alpha_t(f_1\bullet f_2)(\phi) =\left(\frac{H(\cE,h_A,h_B)}{H(\tilde \cE, \tilde h_A, \tilde h_B)}\right)^{it}   
\sum_{\phi=\phi''\bullet \phi'} f_1(\phi')  f_2(\phi'') $$
$$ =\sum_{\phi=\phi''\bullet \phi'}  \left(\frac{H(\cE',h'_A,h'_B) H(\cE'',h''_A,h''_B)}{H(\tilde \cE',\tilde h'_A,\tilde h'_B) H(\tilde \cE'',\tilde h''_A,h''_B)} \right)^{it} f_1(\phi')  f_2(\phi'') =\sum_{\phi=\phi''\bullet \phi'} \alpha_t(f_1)(\phi')  \alpha_t(f_2)(\phi'') . $$

\endproof

\smallskip
\subsection{Partition function and height zeta function}\label{Zeta1Sec}

We can consider subsystems of the algebras with time evolution discussed above, where
the Hamiltonian $\fH$ generating the time evolution has an associated partition function
$Z(\beta)=\Tr(e^{-\beta \fH})$ that defines a height zeta function. To avoid obvious
sources of infinite multiplicities, we can restrict to the subalgebra $\cA^{(1)}_{\cH^L_A(\K)}$ of $\cA^{(1)}_{\cH(\K,Z)}$
of functions supported on the subset $\cH^L(\K,Z)$ of $\cH(\K,Z)$ consisting of $A$-$A$ bimodules 
$(\cE,h^L_A,h^R_A)$ with 
$H^L(\cE,h^L_A,h^R_A)=\prod_\sigma H_{h^L_A,\sigma}(\cE)=1$, for which 
$H(\cE,h^L_A,h^R)=H^R(\cE,h^L_A,h^R)=\prod_\sigma H_{h^R_A,\sigma}(\cE)$. This is indeed a subalgebra
since for bimodules in $\cH(\K,Z)$ we have $H^L_A(\cE\otimes_A \cF,h^L_A,h^R_A)=
H^L_A(\cE,h^{L,'}_A,h^{R,'}_A) H_A^L(\cF,h^{L,''}_A,h^{R,''}_A)$
and both factors are equal to one if both $\cE$ and $\cF$ are in $\cH^L_A(\K)$, so the tensor product
$\cE\otimes_B \cF$ is also in $\cH^L_A(\K)$. Thus, we consider the subsystem $(\cA^{(1)}_{\cH^L_A(\K)},\alpha_t)$
with the same time evolution discussed above. The Hamiltonian generating this time evolution is given by
$$ \fH \, f(\cE,h^L_A,h^R_A) = \log H^R_A(\cE,h^L_A,h^R_A) \, \, f(\cE,h^L_A,h^R_A), $$ 
with $\alpha_t(f) =e^{-itH} \, f \, e^{it H}$, 
for functions $f\in \cA^{(1)}_{\cH^L_A(\K)}$ seen as acting on the Hilbert space $\ell^2(\cH^L_A(\K))$. The
partition function is then of the form
$$ Z(\beta) = \Tr(e^{-\beta \fH}) = \sum_{(\cE,h^L_A,h^R_A)\in \cH^L_A(\K)} H_B(\cE,h^L_A,h^R_A)^{-\beta}, $$
which gives the height zeta function for our definition of height.
The question of whether this height zeta function is convergent for sufficiently large $\beta >0$ 
corresponds then to the question of whether there are finitely many ``points" $(\cE,h_A,h_B)$ of
bounded height $H_B(\cE,h^L_A,h^R_A) \leq \kappa$ (with $H_A(\cE,h^L_A,h^R_A)=1$), 
which corresponds to the property that the spectrum of
the Hamiltonian  $\fH$ has finite multiplicities, and the question of the asymptotic behavior 
of this height bound. We will return to discuss a simpler version of this height zeta function
after introducing a version of the height function based on the Hattori--Stallings rank in \S \ref{HSrankHeightSec}.

\smallskip
\subsection{Limitations of this notion of height}\label{HSrankHeightSec}

The height function as described in the previous subsections provides us
with a viable analog of the corresponding notion used in \cite{Borek1}, which
in turn extends the commutative case where the height is computed in the
form of a volume (or normalized volume). 

\smallskip

As we discussed in the previous subsections, using this notion of
height, however forces us to make some strong assumptions in
order to ensure the multiplicative behavior with respect to the
composition (tensor product) of bimodules. Moreover, it is clear that the
{\em ad hoc} solution that we adopted to circumvent this problem is
specific to the finite dimensional case (hence to noncommutative
arithmetic curves) and does not generalize to higher dimensional
cases. 

\smallskip

Thus, it seems preferable to rethink the notion of height in this
noncommutative context in such a way that the multiplicative
property under composition of bimodules would be naturally 
satisfied, in a way that does not depend on special conditions
on the bimodules, and that extends to the case of algebras that
are not finite dimensional, with Hilbert bimodules of finite type.

\smallskip

The prototype example of an invariant of bimodules of finite
type that satisfies the multiplicative property is the Jones index
(in the case of simple algebras). We will use this as a model for
the properties that a desirable height function should satisfy.  

\smallskip

As a heuristic justification for making this change of perspective,
and proposing a different choice of height function, it is useful
to keep in mind that in the usual arithmetic geometry context
the height function comes in two flavors, a ``logarithmic form"
and an ``exponential form" (see \cite{BoGu}). The latter is the one that gives rise
to the volume description of the height presented in \eqref{BorekH}
and generalized in \eqref{relHE}. The first is usually taken to
be just the logarithm of the exponential one, $\fh=\log H$.
This logarithmic form implies that if $H$ behaves like a volume,
which scales with the size $\lambda$ in the form ${\rm vol}\sim \lambda^{\dim}$,
then the $\fh=\log H$ would be proportional to the dimension. 
What we propose here is not the usual form $\fh=\log H$ of the height,
with $H$ as in \eqref{BorekH}, \eqref{relHE}, but rather the use of
the dimension itself, in the form of the Hattori--Stallings {\em rank element} already
discussed in \S \ref{rankSec}. 

\smallskip

If we define $$\fh_A(\cE):=r_A(\cE)\in HH_0(A),$$ the Hattori--Stallings rank, and 
$$\fh_{A,\sigma}(\cE,h_A,h_B):=\tilde\sigma(r_A(\cE)),$$ with an
embedding $\tilde\sigma$ of $Z(A)$ that extends the 
embedding $\sigma$ of $\K$, then we obtain a better notion 
of height on bimodules that does not require restricting the
category of bimodules as in Definition~\ref{bimodZ}.

\smallskip

A further advantage of working with this definition of height $\fh_A(\cE):=r_A(\cE)$
is that the Hattori--Stallings rank is independent of the generators $\{ x_i \}$ 
of the projective module $\cE$. Indeed, the rank element can be computed
using a choice of $\{ x_i \}$ and of a dual basis $\{ \xi_i \}$ of $\cE^\vee$ by
$r_A(\cE)=\sum_i \xi_i(x_i)$, but the result is independent of the choice of 
$\{ x_i \}$ and $\{ \xi_i \}$. Thus, we do not have to optimize over a choice of the basis,
unlike in the case of the height \eqref{relHE} previously discussed. 

\smallskip

\begin{prop}\label{heightrankdyn}
Consider arithmetic hermitian bimodules $(\cE,h_A,h_B)$ in the category ${}_{(A,\cO_A)}\cH_{(B,\cO_B)}$
with the height defined as
\begin{equation}\label{fhEAB}
\fh(\cE,h_A,h_B)=\prod_\sigma \frac{\sigma(r_B(\cE)}{\sigma(r_A(\cE)},
\end{equation}
where $\sigma$ ranges over the archimedean embeddings. Let $\cA^{(1)}_{\cH(\K)}$ be
the convolution algebra of finitely supported functions over the set of bimodules in ${}_{(A,\cO_A)}\cH_{(B,\cO_B)}$
with the convolution product as in \eqref{convol1}. Then setting $$\alpha_t(f)\,(\cE,h_A,h_B)=
\fh(\cE,h_A,h_B)^{it} \, f(\cE,h_A,h_B)$$
determines a time evolution on $\cA^{(1)}_{\cH(\K)}$.
\end{prop}

\proof As above, we can compute the Hattori--Stallings rank as $r_A(\cE)=\sum_i \xi_i(x_i)$, 
for a choice of generators $\{ x_i \}$ of $\cE$ and a dual set $\{ \xi_i \}$ in $\cE^\vee$. 
In particular, for $\xi_i =h_A(\cdot, x_i)$, we obtain $r_A(\cE)=\sum_i h_A(x_i,x_i)$. 
Thus, we are replacing the volume $\prod_i h_A(x_i,x_i)$ of our previous definition
of height \eqref{relHE} with the trace $\sum_i h_A(x_i,x_i)$. This sum is in $Z(A)=\oplus_i \bL_i$
for some field extensions $\bL_i$ of the number field $\K$. 
Given an archimedean embedding, $\sigma: \K \hookrightarrow \R$ or $\C$, extended
to an embedding $\tilde\sigma$ of the $\bL_i$, we obtain $\tilde\sigma(r_A(\cE))\in \R$ or $\C$.
Since the $h_A(x_i,x_i)$ are positive elements of $A$, the value of $\tilde\sigma(r_A(\cE))$ is 
actually in $\R^*_+$. Thus, we have $\fh(\cE,h_A,h_B)^{it} \in U(1)$ for $t\in \R$ defining
the time evolution. We need to know that $\sigma_t(f_1\star f_2)=\sigma_t(f_1)\star \sigma_t(f_2)$
with respect to the product \eqref{convol1}. This follows by verifying that 
$$ \fh(\cE\otimes_B \cF,h_A,h_C) =\fh(\cE,h'_A,h'_B)\cdot \fh(\cF,h''_B,h''_C), $$
for a tensor product of bimodules $\cE\otimes_B \cF$, with the
hermitian structures $h_A$ and $h_C$ as in Lemma~\ref{tensprodE}. This is indeed
the case because we have 
$$ r_A(\cE\otimes_B \cF)= \sum_{ij} h'_A(x'_i h''_B(x''_j, x''_j), x_i) = \sum_i h'_A(x'_i \cdot r_B(\cF), x'_i) $$
$$ r_C(\cE\otimes_B \cF)= \sum_{\ell.r} h''_C(u''_\ell, h'_B(u'_r,u'_r) u''_\ell)=\sum_\ell h''_C(u''_\ell, r_B(\cE) \cdot 
u''_\ell), $$
where $r_B(\cF)$ and $r_B(\cE)$ are in $Z(B)$. For a fixed embedding $\sigma: \K \hookrightarrow \R$ or $\C$, 
and an embedding $\tilde\sigma$ of the field extensions in $Z(A)$, $Z(B)$, and $Z(C)$, that restricts to $\sigma$
on the subfield $\K$, the hermitian forms $(h'_{A,\sigma}, h'_{B,\sigma})$ on $\cE_\sigma$ and
$(h''_{B,\sigma}, h''_{C,\sigma})$ on $\cF_\sigma$, with values in $A_\sigma$ and $B_\sigma$,
respectively $B_\sigma$ and $C_\sigma$, give 
$$ \sigma(r_A(\cE))=\sum_i h'_{A,\sigma}(x'_i \cdot \sigma(r_B(\cF)), x'_i) =
\sigma(r_B(\cF))\cdot \sigma(r_A(\cE)) , $$
$$ \sigma(r_C(\cE\otimes_B \cF))=\sum_\ell h''_{C,\sigma}(u''_\ell, \sigma(r_B(\cE)) \cdot 
u''_\ell) =\sigma(r_C(\cF))\cdot \sigma(r_B(\cE)), $$
so that we obtain
$$ \fh(\cE\otimes_B \cF,h_A,h_C)=\prod_\sigma \frac{\sigma(r_C(\cF))}{\sigma(r_B(\cF))}    \frac{\sigma(r_B(\cE))}{\sigma(r_A(\cE))}. $$
This completes the proof.
\endproof

\smallskip

The Hattori--Stallings rank is known to agree with the Jones index,
\cite{Kad}, \cite{KaKa}, \cite{Wata}, in the cases we will be considering
in \S \ref{higherdimSec}, so this discussion justifies and introduces
the setting for higher dimensional arithmetic noncommutative spaces
that we will be discussing below in \S \ref{higherdimSec}.

\smallskip
\subsection{Height zeta function with the Hattori--Stallings rank} \label{HSzetaSec}

We can consider again a subsystem with a partition function that provides a
height zeta function, for the case of the height $\fh(\cE,h_A,h_B)$. As we discussed
in \S \ref{Zeta1Sec}, in order to avoid infinite multiplicities in the spectrum of the
Hamiltonian that generates the time evolution, we restrict to the subcategory $\cH^L_A(\K)$ of $\cH(\K)$
$A$-$A$ bimodules $(\cE,h^L_A,h^R_A)$ and with rank (with respect to the structure of left $A$-module)
fixed to be $r^L_A(\cE)=1$. This determines a subalgebra $\cA^{(1)}_{\cH^L_A(\K)}$ of the algebra of
functions on $\cH(\K)$ with the convolution product corresponding to the tensor product of bimodules,
and the Hamiltonian that generates the time evolution, for the algebra $\cA^{(1)}_{\cH^L_A(\K)}$ acting
on the Hilbert space $\ell^1(\cH^L_A(\K))$, is given by
$$ 
\fH \, f(\cE,h^L_A,h^R_A) = \log\left( \prod_\sigma r^R_{A,\sigma} (\cE)\right) \,\,  f(\cE,h^L_A,h^R_A).
$$ 
and with partition function
$$ Z(\beta)=\Tr(e^{-\beta \fH}) = \sum_{(\cE,h^L_A,h^R_A)\in \cH^L_A(\K)} 
\left( \prod_\sigma r^R_{A,\sigma} (\cE)\right)^{-\beta}. $$
The multiplicities of the eigenvalues of the Hamiltonian $\fH$, which determine if the
series defining the partition function is convergent for sufficiently large $\beta >0$, 
count the number of $A$-$A$ bimodules $(\cE,h^L_A,h^R_A)$ with $r_A^L(\cE)=1$
and with fixed $r_A^R(\cE)=r$. For a finite dimensional semisimple algebra
$A=\oplus_i M_{n_i}(D_i)$ with division algebras $D_i$ over $\K$, we can
consider bimodules for the form $D_i \otimes_\K \oplus_j D_j^{n_j N_j}$ such that
$\sum_j n_j N_j =r$, which are counted by the number of these combinations,
the number of solutions $N_j \geq 0$ of the relation above with $r$ and the $n_i$ given.

\medskip
\subsection{Non-archimedean places}\label{nonarchSec}

So far we only considered the archimedean component of the height function,
associated to the archimedean places $\sigma: \K \hookrightarrow \K_\sigma$
of the number field $\K$. The height function also has components associated to
the non-archimedean places. For example, one sees clearly that the usual definition
of the (archimedean) height function for points in projective spaces, 
$H_\sigma(x)=\max_i |x_i|_\sigma$, for $x=(x_1:\cdots :x_n)$ is only well defined
when the non-archimedean places are taken into account in a product over
all valuations, $H(x)=\prod_v \max_i |x_i|_v$, since by the product formula
$H(\lambda x)=H(x)$ for $\lambda\in \bG_m$. This issue does not arise
in an affine setting, but it is still more natural to regard the height as a product
over all archimedean and non-archimedean valuations.

\smallskip

We discuss here briefly how to extend the height functions discussed in the
previous subsection to the non-archimedean places. 

\smallskip

Since we are working in a noncommutative setting, we recall briefly how one
thinks of valuations in the context of finite dimensional algebras. Given a
semisimple finite dimensional algebra $A=\oplus_i M_{n_i}(D_i)$ over a number
field $\K$, let $\K_v$ denote the completions at the non-archimedean valuations.
These are extensions of the p-adic fields $\Q_p$ for primes $p$. We write  
$A_v=A\otimes_\K \K_v$ for the corresponding algebras over $\K_v$. The
fields $\K_v$ are henselian, namely a valuation on $\K_v$ has a unique 
extension to each field algebraic over $\K_v$, \cite{Rib}. Let $D$ be a
division algebra over $\K_v$ and let $Z(D)=\bF$, an extension of $\K_v$.
There are different possible ways of defining valuations on a division
algebra, summarized in \cite{Wad}.  We take here the following definition:
a valuation on a division algebra $D$ is a function $v: D^* \to \Gamma$,
where $\Gamma$ is a totally ordered abelian group such that
$$ v(ab)=v(a)+v(b) \ \ \ \text{ and } \ \ \ v(a+b) \geq \min\{ v(a), v(b) \} \, \text{ for } b\neq -a. $$
Theorem~2.1 of \cite{Wad} shows that, for a division algebra $D$ with $Z(D)=\bF$ and $\dim_\bF(D)<\infty$,
a valuation $v$ on $\bF$ extends (uniquely) to a valuation on $D$ if and only if $v$ extends
uniquely to any field $\bL$ with $\bF \subset \bL \subset D$. In particular, if $\bF$ is henselian,
then a valuation on $\bF$ extends uinquely to a valuation on any finite dimensional
division algebra $D$ with $Z(D)=\bF$. Thus, the non-archimedean valuation $v$
on $\K_v$, extends uniquely to a division algebra over $\K_v$. Moreover, a valuation
on a division algebra $D$ over $\bK_v$ restricts to compatible valuations on subfields of $D$
hence they all arise in this way. Thus, it suffices to consider the non-archimedean
places of $\K$ with the corresponding valuations $v$ and the associated completions $\K_v$
to account for all valuations on division algebras as well.

\smallskip

Consider then arithmetic bimodules $(\cE,h_A,h_B)$ with hermitian structures $h_A, h_B$.
Consider the elements $h_A(x_i,x_i)\in A$ and $h_B(y_j,y_j) \in B$, for sets of generators $\{ x_i \}$
and $\{ y_j \}$. In the case of the height function $H(\cE,h_A,h_B)$ of \eqref{relHE}, and bimodules
$(\cE,h_A,h_B)$ in the class $\cH(\K,Z)$ of Definition~\ref{bimodZ}, we have 
$h_A(x_i,x_i)\in Z(A)=Z$ and $h_B(y_j,y_j) \in Z(B)=Z$ and we can compose 
these elements with the valuation $v$ (the unique extension to the $\bL_i$ of 
the valuation $v$ on $\K_v$). Thus, we can define $H_v(\cE,h_A,h_B)$ as in \eqref{relHE}
in the form 
$$ H_v(\cE,h_A,h_B) := \frac{ \inf_{\{ y_j\} \in \cB_B(\cE)}| \prod_j h_B(y_j,y_j) |_v}{\inf_{\{ x_i\} \in \cB_A(\cE)} | \prod_i h_A(x_i,x_i) |_v}, $$
where a shortest basis is used to evaluate these ``volumes". We then set
$$ H(\cE,h_A,h_B) :=\prod_v H_v(\cE,h_A,h_B) \cdot \prod_\sigma H_\sigma (\cE,h_A,h_B), $$
with $v$ ranging over the non-archimedean places of $\K$ and $\sigma$ over the archimedean
places. Note that because at each place a (different) shortest basis $\{ x_i \}$ is chosen that minimizes the 
corresponding ``volume" $| \prod_i h_A(x_i,x_i) |_v$, the product over places considered above is not subject
to the product formula $\prod_v | \lambda |_v \cdot \prod_\sigma |\lambda|_\sigma =1$ for all
$\lambda\in \bL$ (a number field), for the product over all archimedean and non-archimedean
valuations of $\bL$. 
The time evolutions described in the previous subsections extend to these
height functions that include the contributions of the non-archimedean places.

\smallskip

In the case of the height function defined using the rank element, and for any arithmetic 
hermitian bimodule, however, we cannot just define the height $ \fh(\cE,h_A,h_B)$  in the same way
as above, because for $r_A(\cE)\in \bL$ the product formula would give 
$$ \prod_v | r_A(\cE) |_v \cdot \prod_\sigma | r_A(\cE) |_\sigma = 1. $$
This means that, in one defines the non-archimedean part of the height as in the
previous case, then the non-archimedean part of the height would simply be equivalent to 
the archimedean part by the relation
$$ \fh^{nar}(\cE,h_A,h_B) = \prod_v \frac{| r_B(\cE) |_v}{| r_A(\cE) |_v} =
 \fh^{ar}(\cE,h_A,h_B)^{-1} =\prod_\sigma \frac{| r_A(\cE) |_\sigma}{| r_B(\cE) |_\sigma}, $$
up to a change of direction of the time evolution they induce. Thus, in the case of
the height function defined by the Hattori--Stallings rank, one can restrict to only considering the
archimedean part. 

\smallskip

There is another possible way to extend the Hattori--Stallings rank to
a height function with a non-archimedean components, so that the product over
all archimedean and non-archimedean places is non-trivial. If we view the archimedean part as
$$ \prod_\sigma | r_A(\cE) |_\sigma = \prod_\sigma \sum_i h_{A,\sigma}(x_i,x_i)= \prod_\sigma
\sum_i \| x_i \|_{A,\sigma}^2, $$
then it is natural to define
$$ \fh^{ar}(\cE,h_A,h_B) :=\prod_\sigma \frac{| r_B(\cE) |_\sigma^{1/2}}{| r_A(\cE) |_\sigma^{1/2}}=
\prod_\sigma \frac{(\sum_j \| y_j \|_{B,\sigma}^2)^{1/2}}{(\sum_i \| x_i \|_{A,\sigma}^2)^{1/2}}. $$
By comparison with the usual definition of the height of points $x=(x_1,\ldots, x_n)\in \bar\Q^n$ in the form
$$ H(x)=H^{nar}(x) H^{ar}(x) = \prod_v (\max_i |x_i|_v) \cdot \prod_\sigma (\sum_i |x_i|_\sigma^2)^{1/2}, $$
one can choose to set
$$ \fh(\cE,h_A,h_B)=\fh^{nar}(\cE,h_A,h_B)\fh^{ar}(\cE,h_A,h_B) $$
where the archimedean part is defined as above and 
\begin{equation}\label{rankheightnarch}
\fh^{nar}(\cE,h_A,h_B) := \prod_v \frac{\inf_{\{ y_j \} \in \cB_B(\cE)} \max_j | h_B(y_j,y_j) |_v}{\inf_{\{ x_i \} \in \cB_A(\cE)} \max_i | h_A(x_i,x_i) |_v} . 
\end{equation}
Notice, however, that with this choice of non-archimedean part of the height, we have reintroduced the
problem of optimizing over the choice of a basis and also the problem of the compatibility of the
height with the tensor product of bimodules, which forces us then to restrict the category of bimodules
to the $\cH(\K,Z)$ of Definition~\ref{bimodZ}, both problems that the use of the Hattori--Stallings rank
was meant to avoid. 

\smallskip

Thus, in the following, in the case of the height based in the Hattori--Stallings rank, we will simply 
restrict to the archimedean part.

\medskip
\section{More general algebras}\label{higherdimSec}

In the previous section we have taken inspiration from the setting of \cite{Borek1},
designed to provide a noncommutative geometry setting for the Arakelov geometry
of noncommutative arithmetic curves. When we try to extend these ideas beyond
the case of zero-dimensional varieties over a number field $\K$ (hence one-dimensional
varieties over ${\rm Spec}(\cO_\K)$) and their noncommutative counterparts, 
we encounter a more serious problem in how to introduce the data at infinity,
that is, the contribution of the archimedean places. In the case of arithmetic
surfaces (curves over a number field $\K$, seen as surfaces over ${\rm Spec}(\cO_\K)$),
noncommutative geometry was used in different ways: in \cite{ConsMar1}, \cite{ConsMar2},
\cite{Mar-book} noncommutative geometry was used to model the fiber at
infinity of an arithmetic surface, following a model developed in \cite{Man3} based
on Schottky uniformization and geodesics in a hyperbolic handlebody (see also \cite{ManMar}
for a physical interpretation in the setting of AdS/CFT holography).  In \cite{Borek2} 
noncommutative geometry is used in a different way, by developing noncommutative 
versions of arithmetic surfaces, based on noncommutative projective schemes.
In the context of Arakelov geometry it is natural to work with projective schemes, because
one is interested primarily in developing a good intersection theory. However, from the
point of view of noncommutative geometry this creates complications in relating the
algebro-geometric noncommutative projective schemes to analytic noncommutative
spaces that should provide the data at the archimedean places. While this problem
is solved in the case of noncommutative arithmetic surfaces in \cite{Borek2} by a
suitable definition of hermitian bundles, we take here a different viewpoint on
noncommutative arithmetic spaces, which is based on affine geometry. This
is sufficient for our purposes, and it will have the advantage that it makes more
direct and transparent the reole of the archimedean places. 

\smallskip
\subsection{Arithmetic noncommutative spaces}

For our purposes, we take the following definition of an arithmetic noncommutative space.

\begin{defn}\label{arithmNCvar} {\rm 
Let $\K$ be a number field and $\cO_\K$ its ring of integers. 
An arithmetic noncommutative space in arbitrary dimension is given by the following data.
\begin{enumerate}
\item $A$ is a finitely generated associative algebra over $\K$.
\item $\cO_A$ is a subring of $A$ with the structure of flat $\cO_\K$-module.
\item As $\K$-vector spaces $A=\cO_A\otimes_{\cO_\K} \K$. 
\item For every archimedean place $\sigma: \K \hookrightarrow \K_\sigma$
with $\K_\sigma=\R$ or $\C$, the real/complex algebra $A_\sigma=A\otimes_{\K,\sigma}\K_\sigma$
has an involution $\star$.
\item $A_\sigma$, for $\sigma$ complex, and $A_\sigma\otimes_\R \C$ for $\sigma$ real, are 
dense involutive subalgebras of a complex $C^*$-algebra $\bar A_\sigma$. 
\item There is a normalized trace $\tau_\sigma: \bar A_\sigma \to \K_\sigma$ such that
$\tau_\sigma |_{\cO_A}$ takes values in $\cO_\K$.
\end{enumerate}   }
\end{defn}

\smallskip

This notion of arithmetic noncommutative spaces only involves Type II algebras. Thus, for
example, certain noncommutative spaces of arithmetic relevance such as the endomotives of
\cite{CCM}, including the Bost--Connes type algebras over number fields of \cite{Yalk}, are
not included in this definition. While such algebras certainly should be regarded as arithmetic spaces,
they would not be of finite type over ${\rm Spec}(\cO_\K)$, while our
notion is meant to capture this finite type condition. 

\smallskip

\begin{ex}\label{arNCTori}{\rm The arithmetic noncommutative torus is the finitely generated
algebra $A(\bT_\theta)$ over $\Q$
with generators $U,V,U^*,V^*$ and relations $U^*U=UU^*=1$, $V^*V=VV^*=1$ and $VU=e^{2\pi i\theta}UV$
for some fixed $\theta\in \R$. The subring $\cO_{A(\bT_\theta)}$ is generated over $\Z$ by the same generators
and relations. As a $\Z$-module it is the span $\Z\langle U^n V^m\,|\, n,m\in \Z \rangle$, which is an abelian group
with no torsion, hence a flat $\Z$-module. For the unique real embedding of $\Q$ the algebra
$A(\bT_\theta)_\sigma\otimes_\R \C$ is a dense subalgebra of the usual $C^*$-algebra of 
the noncommutative torus. The trace $\tau(U^n V^m)=\delta_{n,0} \delta_{m,0}$ maps $\cO_{A(\bT_\theta)}$
to $\Z$. }\end{ex}

\smallskip

For two $C^*$-algebras $\bar A$ and $\bar B$, we recall the notion of 
Hilbert $\bar A-\bar B$ bimodule, \cite{KajWat}. 

\begin{defn}\label{Hilbertbimod}{\rm
A Hilbert $\bar A-\bar B$ bimodule $H$ of finite type is defined by the following properties:
\begin{itemize}
\item $H$ is an $\bar A- \bar B$-bimodule, which is finitely generated projective as a right $\bar B$-module
and as a left $\bar A$-module
\item $H$ is self-dual, with respect to the duality given by the conjugate bimodule.
\item There are an $\bar A$-valued inner product $_{\bar A}\langle\cdot, \cdot \rangle$
and a $\bar B$-valued inner product $\langle\cdot, \cdot \rangle_{\bar B}$, with the first
left-linear and right-conjugate-linear and the second left-conjugate-linear and right-linear.
\item For all $a\in \bar A$, $b\in \bar B$, and $x,y\in H$ 
$$ {}_{\bar A}\langle ax,y \rangle = a\, {}_{\bar A}\langle x,y \rangle, \ \ \ \text{ and } \ \ \ 
{}_{\bar A}\langle x,ay \rangle = {}_{\bar A}\langle x,y \rangle\, a^* $$
$$ \langle x,yb \rangle_{\bar B}= \langle x,y \rangle_{\bar B}\, b \ \ \  \text{ and } \ \ \
\langle xb,y \rangle_{\bar B} = b^*\, \langle x,y \rangle_{\bar B} $$
\item For all $x\in H$, $_{\bar A}\langle x,x \rangle\geq 0$ with $_{\bar A}\langle x,x \rangle=0$
iff $x=0$ and $\langle x,x \rangle_{\bar B}\geq 0$ with $\langle x,x \rangle_{\bar B}=0$ iff $x=0$.
\item For all $x,y\in H$, 
$$ {}_{\bar A}\langle x,y \rangle = {}_{\bar A}\langle y,x \rangle^* \ \ \  \text{ and } \ \ \
\langle x,y \rangle_{\bar B} = \langle y,x \rangle_{\bar B}^* $$
\item The norms $\| x \|_A:=\| {}_{\bar A}\langle x,x \rangle \|_{\bar A}^{1/2}$ and
$\| x\|_B:=\| \langle x,x \rangle_{\bar B} \|_{\bar B}^{1/2}$ satisfy an estimate
$$ C_1 \| x \|_A \leq \| x\|_B \leq C_2 \| x \|_A $$
for some constants $C_1,C_2>0$ and for all $x\in H$.
\item $H$ is complete with respect to the $\|\cdot \|_A$ norm (equivalently, the $\| \cdot \|_B$
norm). 
\end{itemize}
}\end{defn} 

\smallskip

\begin{rem}\label{fintype}{\rm The finite type condition, that $H$ is finitely generated projective both
as a right $\bar B$-module and as a left $\bar A$-module, and that it is self-dual with respect to
conjugation is equivalent to the condition that there are a left $\bar A$-basis $\{ v_1, \ldots, v_m \}$
and a right $\bar B$-basis $\{ u_1, \ldots, u_n \}$ such that, for all $x\in H$,
\begin{equation}\label{LRbasis}
x=\sum_{i=1}^m {}_{\bar A}\langle x,v_i\rangle v_i   \ \ \ \text{ and } \ \ \
x=\sum_{j=1}^n u_j \,\langle u_j, x \rangle_{\bar B}.   
\end{equation}
See Lemma~1.7 of \cite{KajWat},
and see also the comment on duality on p.~3443 of \cite{KajWat}.
}\end{rem}

\smallskip

In terms of a right $\bar B$-basis $\{ u_1, \ldots, u_n \}$, the structure of $H$ as a finitely
generated projective right $\bar B$-module can be seen explicitly by identifying $H\simeq p \bar B^n$,
where $p\in M_n(\bar B)$ is a projection given by $p_{ij}=\langle u_i, u_j \rangle_{\bar B}$.
The maps from $H$ to $p \bar B^n$ and viceversa are given by $x\mapsto (\langle u_i,x \rangle_{\bar B})_i$,
which maps $H$ to the range of $p$ in $\bar B^n$, and its inverse map
$(y_i)\mapsto \sum_i u_i y_i$ from $p \bar B^n$ to $H$. (See Lemma~1.11 of \cite{KajWat}.)
Also note that the condition on the equivalence of the norms $\| x \|_A$ and $\| x\|_B$ in
Definition~\ref{Hilbertbimod} and of completeness in these norms follow automatically 
from the finite type condition, as in Remark~\ref{fintype} (see Lemma~1.11 and 
Proposition~1.18 of \cite{KajWat}.)

\smallskip

We can then generalize the definition of hermitian bimodules given in Definition~\ref{Hbimods}
in the following way.

\begin{defn}\label{HbimodsGen} {\rm  Let $(A,\cO_A)$ and $(B,\cO_B)$ be
arithmetic noncommutative spaces as in Definition~\ref{arithmNCvar}. 
The category $_{(A,\cO_A)}\cH_{(B,\cO_B)}$ of arithmetic hermitian bimodules of finite type 
has objects $(\cE,h_A,h_B)$ satisfying all the properties of Definition~\ref{Hbimods}
as well as the following:
\begin{itemize}
\item $\cE$ is an $\cO_A-\cO_B$ bimodule, 
and a left-right $\cO_\K$-lattice, such that $A\otimes_{\cO_A} \cE$ is
finitely generated projective as an $A$-module and $\cE\otimes_{\cO_B} B$ is 
finitely generated projective as $B$-module.
\item for any archimedean place $\sigma$ of $\K$ the
bimodule $\cE_\sigma:=A_\sigma \otimes_{\cO_A} 
\cE \otimes_{\cO_B} B_\sigma$ is endowed with a pair $h=(h_A,h_B)$ 
with $h_A: \cE_\sigma \times \cE_\sigma \to A_\sigma$
and $h_B: \cE_\sigma \times \cE_\sigma \to B_\sigma$
\item $h_A$ is left-linear and right-conjugate-linear and defines an $A_\sigma$-valued inner product
on $\cE_\sigma$. 
\item $h_B$ is left-conjugate-linear and right-linear and defines a $B_\sigma$-valued inner product
on $\cE_\sigma$. 
\item There is a Hilbert $\bar A_\sigma - \bar B_\sigma$ bimodule of finite type $\bar \cE_\sigma$ 
with ${}_{\bar A_\sigma}\langle x,y\rangle=h_A(x,y)$ and $\langle x,y\rangle_{\bar B_\sigma}=h_B(x,y)$
\item $\bar \cE_\sigma$ 
is the completion of $\cE_\sigma$ in the norm $\| x \|_A=\| h_A(x,x) \|_{\bar A_\sigma}^{1/2}$
(equivalently, in the norm $\| x \|_B=\| h_B(x,x) \|_{\bar B_\sigma}^{1/2}$). 
\end{itemize}
Morphisms $\phi: (\cE,h_A,h_B)\to (\cE', h_A', h_B')$ are morphisms $\phi:\cE\to \cE'$ 
of $\cO_A-\cO_B$ bimodules such that the induced map $\phi_\sigma: \bar\cE_\sigma\to \bar\cE'_\sigma$
is a finite rank operator that preserves the hermitian structures,
$h_A'(\phi_\sigma(x),\phi_\sigma(y))=h_A(x,y)$ and $h_B'(\phi_\sigma(x),\phi_\sigma(y))=h_B(x,y)$.
}\end{defn}

\smallskip

With respect to a right $\bar B$-basis $\{ u_1, \ldots, u_n \}$ of $\bar \cE_\sigma$
a bimodule homomorphism $\phi_\sigma: \bar\cE_\sigma\to \bar\cE'_\sigma$ satisfies 
$$ \phi_\sigma(x)=\sum_{i=1}^n T(u_i) \, \langle u_i, x\rangle_{\bar B_\sigma}, $$
and similarly for a left $\bar A$-basis $\{ v_1, \ldots, v_m \}$ of $\bar \cE_\sigma$.
Thus, for the condition that morphisms preserve the hermitian structure it suffices
that on the basis elements $h_B(\phi_\sigma(u_i),\phi_\sigma(u_j))=h_B(u_i,u_j)$
and $h_A(\phi_\sigma(v_i),\phi_\sigma(v_j))=h_A(v_i,v_j)$.

\smallskip

\begin{rem}\label{functpts2}{\rm
The functor of points of an arithmetic noncommutative space $(A,\cO_A)$ 
assigns to $(B,\cO_B)$
the category $_{(A,\cO_A)}\cH_{(B,\cO_B)}$ of arithmetic hermitian bimodules of finite type
as the $(B,\cO_B)$ points of $(A,\cO_A)$, and to $(B,\cO_B)$-$(B',\cO'_B)$ bimodules $\cF$
the functor $\cF: {}_{(A,\cO_A)}\cH_{(B,\cO_B)}\to {}_{(A,\cO_A)}\cH_{(B',\cO'_B)}$ given by
tensoring with $\cF$.
}\end{rem}

\smallskip
\smallskip
\subsection{The case of noncommutative tori}

We consider again the example of arithmetic structures on noncommutative
tori, mentioned in Example~\ref{arNCTori} above.

\smallskip

The Rieffel construction of finite projective modules on noncommutative tori \cite{Rief}, \cite{Rief2} 
is based on using imprimitivity bimodules. These are Hilbert bimodules ${}_A\cF_B$, for unital
$C^*$-algebras $A,B$, with
the property that the $C^*$-algebra valued inner products satisfy the relation
\begin{equation}\label{imprimrel}
\langle x, y \rangle_A \, z = x \, {}_B\langle y, z \rangle
\end{equation}
for all $x,y,z\in {}_A\cF_B$. An imprimitivity bimodule implements a strong Morita equivalence
between the  $C^*$-algebras $A$ and $B$, \cite{Rief3}. Thus, these are regarded as a good notion of
``isomorphism" of the corresponding non-commutative spaces, in a setting where morphisms
are defined by (some class of) Hilbert bimodules. The construction of projective modules
is based on the observation that, if ${}_A\cF_B$ is an imprimitivity bimodule and $p$ is a 
projection in $A$, then $p\cF$ is a projective $B$-module, since there will be two sets 
$\{ x_i \}_{i=1}^n$ and $\{ y_i \}_{i=1}^n$ of elements of $\cF$ such that
$$ \sum_i \langle x_i, y_i \rangle_A = p. $$
Then for each $z\in p\cF$ one has $z=pz=\sum_i \langle x_i, y_i\rangle_A z = \sum_i x_i {}_B\langle y_i,z \rangle$
so that the $x_i$ give a finite set of generators for $p\cF$ and the maps
\begin{equation}\label{PhiPsimaps}
\begin{array}{rcl}
 \Psi: B^n \to p\cF & \text{ and } & \Phi: p\cF \to B^n \\[2mm]
 \Psi: (b_i) \mapsto \sum_i x_i b_i  & \text{ and } & \Phi: z \mapsto ({}_B\langle y_i, z\rangle) 
 \end{array}
 \end{equation}
identify $p\cF$ with a direct summand of the free module $B^n$, so that $p\cF$ is indeed 
a projective $B$-module, see Proposition~1.2 of \cite{Rief2}. 
Thus, the construction of projective $B$-modules follows from
the construction of projections $p$ in the Morita-equivalent algebra $A$. 

\smallskip

In particular, an arithmetic structure on a projective module $X =p \cF$ over a noncommutative
torus $B=A(\bT_\theta)$ is obtained by considering the image
$\Psi(\cO_{A(\bT_\theta)}) \subset p\cF$,
\begin{equation}\label{pEarithm}
\begin{array}{rl} 
 \Psi(\cO_{A(\bT_\theta)})=& \{ \xi \in p\cF\,|\, \xi =\sum_i x_i b_i , \text{ with } b_i \in \cO_{A(\bT_\theta)}\} \\[2mm]
=& \{ \xi\in p\cF \,|\, {}_{A(\bT_\theta)}\langle y_i, \xi \rangle \in \cO_{A(\bT_\theta)} \} . 
\end{array}
\end{equation}

\smallskip

The construction of projections in a noncommutative torus $A=A(\bT_\theta)$
is obtained as in \cite{Rief} by considering the $A$-valued inner product
$\langle \xi, \xi \rangle_A$ of elements $\xi\in {}_A\cF_B$. The resulting
$p_\xi=\langle \xi, \xi \rangle_A$ is a projection if and only if it satisfies the identity
$ \xi \, {}_B\langle \xi, \xi \rangle = \xi $.
Indeed, one can see that
$$ p_\xi^2 =\langle \xi, \xi \rangle_A \, \langle \xi, \xi \rangle_A =
\langle \langle \xi, \xi \rangle_A \, \xi, \xi \rangle_A= \langle \xi \, {}_B\langle \xi, \xi \rangle, \xi \rangle_A=
\langle \xi, \xi \rangle_A=p_\xi $$
and $p_\xi^*=p_\xi$. The converse follows similarly (see Lemma~3.2 of \cite{Luef}).
There are several explicit constructions of projections in noncommutative tori. In particular, 
Boca \cite{Boca} presented a construction based on theta functions, and Luef and Manin
\cite{LuMa} and Luef \cite{Luef} gave a general construction based on Gabor frames. 

\smallskip

One can then obtain hermitian bimodules of finite type in the following way.
Here we assume we have fixed a specific construction that associates to a pair of Morita
equivalent $C^*$-algebras of irrational noncommutative tori and an imprimitivity bimodule
implementing the Morita equivalence a projection in one of the two $C^*$-algebras obtained
as described above. 

\begin{prop}\label{bimodNCtori}
Let $A_\theta=A(\bT_\theta)$ and $A_{\theta'}=A(\bT_{\theta'})$ be the $C^*$-algebras
of two irrational noncommutative tori. Let $\tilde\theta$ and $\tilde\theta'$ be a choice
of points in the $\GL_2(\Z)$ orbits of $\theta$ and $\theta'$ respectively, under
the action on $\P^1(\R)$ by fractional linear transformations. Such a choice
determines an $(A_\theta,\cO_{A_\theta})$--$(A_{\theta'},\cO_{A_{\theta'}})$
hermitian bimodule as in Definition~\ref{HbimodsGen}.
\end{prop}

\proof Since $\theta$ and $\tilde\theta$ (respectively, $\theta'$ and $\tilde\theta'$)
are in the same $\GL_2(\Z)$ orbit by fractional linear transformations, the $C^*$-algebras
$A_\theta$ and $A_{\tilde\theta}$ (respectively, $A_{\theta'}$ and $A_{\tilde\theta'}$) are
Morita equivalent, with the equivalence implemented by imprimitivity bimodules 
${}_{A_\theta}\cF_{A_{\tilde\theta}}$ and ${}_{A_{\tilde\theta'}}\cF'_{A_{\theta'}}$. 
Let $p_{\tilde\theta}$ and $p_{\tilde\theta'}$
be projections in $A_{\tilde\theta}$ and $A_{\tilde\theta'}$, respectively, and let $X=\cF\, p_{\tilde\theta}$
and $X'=p_{\tilde\theta'}\cF'$ be the corresponding projective modules, constructed as recalled above.
Consider then the $A_\theta$--$A_{\theta'}$ Hilbert bimodule $\cE =X\otimes_\C X'$. As above,
let $\{ x_i \}$ and $\{ y_i \}$ be elements of $\cF$ and $\{ x_j' \}$ and $\{ y_j'\} $ be elements of $\cF'$
such that 
\begin{equation}\label{xiyip}
 \sum_i {}_{A_{\tilde\theta}}\langle y_i, x_i \rangle = p_{\tilde\theta} \ \ \ \text{ and } \ \ \ 
\sum_j \langle x_j', y_j' \rangle_{A_{\tilde\theta'}} = p_{\tilde\theta'}. 
\end{equation}
Let $\Phi_{\tilde\theta}, \Psi_{\tilde\theta}$ and $\Phi_{\tilde\theta'}, \Psi_{\tilde\theta'}$ be the
maps defined as in \eqref{PhiPsimaps} with $\Psi_{\tilde\theta}: A_{\theta}^n\to X$,
$\Psi_{\tilde\theta}(a_i)=\sum_i a_i x_i$ and $\Psi_{\tilde\theta'}: A_{\theta'}^n\to X'$,
$\Psi_{\tilde\theta'}(a_j')=\sum_j x'_j a'_j$. Elements $\xi\otimes\xi'\in \cE$ can be
written as $\xi\otimes \xi' =\sum_{i,j} \langle \xi, y_i \rangle_{A_\theta}\, x_i \otimes x_j' \,
{}_{A_{\theta'}}\langle y_j',\xi \rangle = \sum_{i,j} a_i\, x_i\otimes x'_j \, a'_j$.
We define the arithmetic submodule $\cE_\Z$ as  in \eqref{pEarithm} by setting
$$ \cE_\Z:=\{ \xi\otimes \xi' =  \sum_{i,j} a_i\, x_i\otimes x'_j \, a'_j \,|\, a_i \in \cO_{A_\theta}, 
a_j' \in \cO_{A_{\theta'}}, \, \forall i,j  \}. $$
We define hermitian structures $h_\theta$ and $h_{\theta'}$ on $\cE_\Z$ as
$h_\theta(\xi,\eta)={}_{A_\theta}\langle \xi, \eta\rangle$ and 
$h_{\theta'}(\xi,\eta)=\langle \xi,\eta\rangle_{A_{\theta'}}$.
\endproof

\smallskip
\subsection{Height function as volume}\label{VolHtypeIISec}

We would like then to assign a height to the bimodules $(\cE, h_A, h_B)$ 
constructed as above. We first discuss how to obtain a version of the height $H(\cE, h_A, h_B)$ 
that corresponds to the expressions \eqref{BorekH} and \eqref{relHE}. We then discuss
how to obtain a notion of height $\fh(\cE, h_A, h_B)$ that generalized the one based
on the Hattori--Stallings rank that we discussed in the previous section. 

\smallskip

We want to assign to an arithmetic hermitian bimodule of finite type 
$(\cE, h_A, h_B)$ a height function of the form
\begin{equation}\label{heightNCtorus}
H(\cE, h_A, h_B)= \frac{{\rm vol}_{B,h_B}(\cE)}{{\rm vol}_{A,h_A}(\cE)}, 
\end{equation}
as a direct analog of \eqref{relHE}. 

\smallskip

In order to obtain the volumes ${\rm vol}_{A,h_A}(\cE)$ and
${\rm vol}_{B,h_B}(\cE)$ we proceed as in the previous section.
Let $(\cE,h_A,h_B)$ be an arithmetic hermitian bimodule of finite type as in Definition~\ref{HbimodsGen}.
Let $\cB_{A_\sigma}(\cE)$ and $\cB_{B_\sigma}(\cE)$ be, respectively, 
the sets of left $\bar A_\sigma$-bases of $\cE_\sigma$
$\{ v_1, \ldots, v_m \}$, with that $v_j\in \cE$, and the set of 
right $\bar B_\sigma$-bases $\{ u_1,\ldots, u_n \}$ of $\cE_\sigma$ with $u_i\in \cE$.

\smallskip

We define the volume via an optimization over the choice of basis. Namely, we set 
$$ {\rm vol}_{A,h_A}(\cE):=\prod_\sigma \inf_{\cB_{A_\sigma}(\cE)} \prod_j \tau_{A_\sigma}(h_A(v_j, v_j)), $$
$$ {\rm vol}_{B,,h_B}(\cE):=\prod_\sigma \inf_{\cB_{B_\sigma}(\cE)} \prod_i \tau_{B_\sigma}(h_B(u_i,u_i)), $$
where $\tau_{A_\sigma},\tau_{B_\sigma}$ denote the unique trace on the algebras $A_\sigma$, $B_\sigma$.

\smallskip

This provides an analog of  \eqref{relHE} extending the notion of height derived from \eqref{BorekH}.
However, as in the case of  \eqref{relHE}, the height function obtained in this way is not well
behaved with respect to taking tensor products of bimodules, hence it is not directly compatible
with the categorical composition operation on our noncommutative notion of ``points". 
To avoid this problem, we consider, as in the previous section, a different notion of height
based on the appropriate notion of ``dimension", instead of a height function based on volumes. 

\smallskip
\subsection{Jones index as a height function}

We return here to the general setting of Definition~\ref{Hilbertbimod}. The case of noncommutative tori
discussed above is included as a special case. 

\smallskip

Let $(\cE,h_A,h_B)$ be a arithmetic hermitian bimodule of finite type as in Definition~\ref{HbimodsGen}
and let $\{ u_1,\ldots, u_n \}$ and $\{ v_1, \ldots, v_m \}$ be, respectively, a right $\bar B_\sigma$-basis 
and a left $\bar A_\sigma$-basis. The right and left index are defined as the elements 
\begin{equation}\label{Jindex}
{\rm Ind}_{A,\sigma}(\cE):=\sum_i h_A(u_i,u_i) \in Z(\bar A_\sigma) \ \ \ \text{ and } \ \ \ 
{\rm Ind}_{B,\sigma}(\cE):=\sum_j h_B(v_j,v_j) \in Z(\bar B_\sigma) 
\end{equation}
with $Z(\bar A_\sigma)$ and $Z(\bar B_\sigma)$ the centers of the respective $C^*$-algebras.
These elements are independent of the choice of basis (Proposition~1.13 of \cite{KajWat}).  
We use here the notation ${\rm Ind}_{A,\sigma}(\cE)$ and ${\rm Ind}_{B,\sigma}(\cE)$ for what would be,
respectively,   
${\rm Ind}^R(\cE_\sigma)$ and ${\rm Ind}^L(\cE_\sigma)$ in the notation of \cite{KajWat}. 

\smallskip

Consider algebras $A,B$ such that $Z(A_\sigma)=Z(B_\sigma)=\K_\sigma$ for all archimedean 
places $\sigma$. Then the Jones index is multiplicative over tensor product of bimodules
$$ {\rm Ind}_{A,sigma}(\cE \otimes_B \cF)={\rm Ind}_{A,sigma}(\cE) \cdot {\rm Ind}_{B,\sigma}(\cF), \ \ \
{\rm Ind}_{C,\sigma}(\cE \otimes_B \cF)={\rm Ind}_{B,\sigma}(\cE) \cdot {\rm Ind}_{C,\sigma}(\cF), $$
see Proposition~1.30 of \cite{KajWat}. 

\smallskip

For algebras with $Z(A_\sigma)=Z(B_\sigma)=\K_\sigma$ one can then define a height
function as in \eqref{fhEAB}
\begin{equation}\label{fhJones}
\fh(\cE,h_A,h_B):= \prod_\sigma \frac{{\rm Ind}_{B,\sigma}(\cE)}{{\rm Ind}_{A,\sigma}(\cE)}.
\end{equation}
Imprimitivity bimodules $(\cE,h_A,h_B)$ that implement a Morita equivalence between $A_\sigma$
and $B_\sigma$, have  ${\rm Ind}_{B,\sigma}(\cE)=1={\rm Ind}_{A,\sigma}(\cE)$, see Corollary~1.19
of \cite{KajWat}. Thus, imprimitivity bimodules are fixed points of the time evolution generated
by the height function \eqref{fhJones} as in Proposition~\ref{heightrankdyn}. 

\smallskip

The height zeta function associated to the height function \eqref{fhJones} as discussed
in \S \ref{HSzetaSec} involves bimodules with fixed values of the Jones index. This
shows that the problem of studying ``points of bounded height" in this setting is potentially
an interesting question with several possible connections to other areas of mathematics. 
In the von Neumann algebra setting, the related question of classifying subfactors of 
fixed index is solved for index less than $4$ (see \cite{Ocn})
where the values of the Jones index are quantized, and also for index between $4$ and $5$ (see \cite{JoMoSny})
where after excluding an infinite family, one can again reduce the question to discrete data,
while the question becomes intractable for higher values of the index for the reasons 
explained in \cite{JoMoSny}.

\bigskip

\subsection*{Acknowledgment} The second author is partially supported by
NSF grant DMS-1707882, and by NSERC Discovery Grant RGPIN-2018-04937 
and Accelerator Supplement grant RGPAS-2018-522593. 


\end{document}